\documentclass[acmlarge,screen,nonacm]{acmart}

\usepackage{pifont}
\usepackage{adjustbox}
\usepackage[utf8]{inputenc}
\AtBeginDocument{%
  }

\setcopyright{none}
\renewcommand\footnotetextcopyrightpermission[1]{}

\usepackage{booktabs,tabularx}
\usepackage{graphicx}   
\usepackage{caption}    
\usepackage{subcaption} 
\usepackage{pdfpages}
\usepackage{pifont}
\usepackage{xcolor}
\usepackage{multirow}
\usepackage{siunitx}
\usepackage[table,xcdraw]{xcolor}  
\usepackage{amsmath}  
\usepackage{tikz}
\usetikzlibrary{arrows.meta,positioning,fit,calc,shadows.blur}

\newcommand{\cmark}{\textcolor{green!60!black}{\ding{51}}} 
\newcommand{\xmark}{\textcolor{red!70!black}{\ding{55}}}   

\newcommand*\systemname{{\bf SIGMA-ASL}}

\usepackage{xcolor}

\newcommand{\added}[1]{#1}

\newif\ifshowrev
\showrevfalse 

\ifshowrev
  \newcommand{\myadded}[1]{\textcolor{blue}{#1}}
\else
  \newcommand{\myadded}[1]{#1}
\fi

\begin{document}

\title{SIGMA-ASL: Sensor-Integrated Multimodal Dataset for Sign Language Recognition}


\author{Xiaofang Xiao}
\orcid{0009-0000-7715-2393}
\affiliation{%
  \institution{School of Software, Shandong University}
  \city{Jinan}
  \country{China}
}
\email{xiaofang.xiao@mail.sdu.edu.cn}

\author{Guangchao Li}
\orcid{0009-0004-9625-1265}
\affiliation{%
  \institution{School of Software, Shandong University}
  \city{Jinan}
  \country{China}
}

\author{Guangrong Zhao}
\orcid{0000-0002-4703-9397}
\affiliation{%
  \institution{School of Software, Shandong University}
  \city{Jinan}
  \country{China}
}

\author{Qi Lin}
\orcid{0000-0003-3676-9789}
\affiliation{%
  \institution{School of Artificial Intelligence, Shandong University}
  \city{Jinan}
  \country{China}
}

\author{Wen Ma}
\orcid{0000-0002-6601-1090}
\affiliation{%
  \institution{School of Foreign Language and Literature, Shandong University}
  \city{Jinan}
  \country{China}
}

\author{Hongkai Wen}
\orcid{0000-0003-1159-090X}
\affiliation{%
  \institution{University of Warwick}
  \city{Coventry}
  \country{United Kingdom}
}

\author{Yanxiang Wang}
\authornote{Corresponding author.}
\orcid{0000-0002-1466-4006}
\affiliation{%
  \institution{School of Software, Shandong University}
  \city{Jinan}
  \country{China}
}

\author{Yiran Shen}
\orcid{0000-0003-1385-1480}
\affiliation{%
  \institution{School of Software, Shandong University}
  \city{Jinan}
  \country{China}
}
\renewcommand{\shortauthors}{Xiao et al.}








\begin{abstract}
Automatic sign language recognition (SLR) has become a key enabler of inclusive human–computer interaction, fostering seamless communication between deaf individuals and hearing communities. Despite significant advances in multimodal learning, existing SLR research remains dominated by vision-based datasets, which are limited by sensitivity to lighting and occlusion, privacy concerns, and a lack of cross-modal diversity. To address these challenges, we introduce \systemname{}, a large-scale multimodal dataset for SLR. The dataset integrates an Azure Kinect RGB-D camera, a millimeter-wave (mmWave) radar, and two wrist-worn inertial measurement units (IMUs) to capture complementary visual, radio-reflection, and kinematic information. Collected in a controlled studio environment with 20 participants performing 160 common American sign language (ASL) signs, \systemname{} provides \added{93,545} temporally synchronized word-level multimodal clips. A unified sensing framework achieves millisecond-level alignment across modalities, enabling reliable sensor fusion and cross-modal learning. We further design standardized preprocessing pipelines and benchmarking protocols under both user-dependent and user-independent settings, offering a comprehensive foundation for evaluating single and multimodal SLR. Extensive experiments validate the dataset’s quality and demonstrate its potential as a valuable resource for developing robust, privacy-preserving, and ubiquitous sign language recognition systems.
\end{abstract}





\maketitle

\section{Introduction}

Sign language is the primary means of daily communication for tens of millions of deaf and hard-of-hearing individuals worldwide~\cite{WFD2023}. Unlike spoken languages that rely on acoustic signals, sign languages convey meaning through coordinated combinations of hand gestures, body postures, and facial expressions. This highly visual and spatial nature enables expressive and efficient communication but also introduces substantial challenges for automatic understanding and recognition, as it involves complex spatio-temporal dynamics, simultaneous multi-joint articulation, and individual variation in signing style.

With the rapid advancement of machine learning, particularly large language models (LLMs), vision–language models (VLMs), and edge computing technologies~\cite{djebrouni2024bias,elhattab2024pastel,liu2020pmc,radford2021learning,li2023blip}, automatic sign language recognition (SLR) has become increasingly feasible, attracting growing research attention~\cite{sarhan2023unraveling,koller2020quantitative,li2020word}. Within the ubiquitous and pervasive computing community, sign language recognition represents an important and emerging research frontier: it exemplifies intelligent human–environment interaction, demands real-time multimodal sensing and interpretation, and connects deeply with the community’s long-term vision of accessible, inclusive, and context-aware computing. Advancing SLR not only promotes accessibility for the deaf community but also drives innovation in multimodal perception, on-device learning, and privacy-preserving interaction, which are key themes in ubiquitous computing research.

SLR can be broadly categorized into isolated and continuous recognition tasks. Isolated SLR (ISLR) focuses on recognizing individual signs or words performed independently, where the temporal boundaries of each sign are clearly defined \cite{huang2018video,alyami2024reviewing,agrawal2016survey}. This formulation simplifies modeling by avoiding co-articulation effects between consecutive signs and is often used in early-stage research and dataset collection. In contrast, continuous SLR (CSLR) aims to recognize entire sentences or phrases comprising a sequence of signs, requiring temporal segmentation, boundary detection, and language modeling to handle transitions and grammatical structures \cite{koller2016deep,koller2020quantitative,camgoz2018neural,cui2019deep,li2023signring}. While CSLR more closely reflects natural communication, ISLR remains fundamental for benchmarking multimodal sensing, building robust lexical models, and enabling scalable annotation and analysis.

Most existing SLR research primarily relies on visual sensing, such as RGB or RGB-D cameras, to capture a signer’s actions and infer the corresponding linguistic meaning~\cite{koller2016deep,zhou2023unified,li2020word}. Vision-based SLR has become a major research focus in the computer vision community, driven by advances in deep learning and the availability of large-scale video datasets~\cite{simonyan2014two,carreira2017quo}. Prominent examples include WLASL~\cite{li2020word}, which provides thousands of word-level glosses, and RWTH-PHOENIX-Weather~\cite{forster2012rwth}, a benchmark for continuous recognition and translation in broadcast environments. Datasets such as CSL500~\cite{huang2018attention} and PopSign ASL~\cite{starner2023popsign} further extend vocabulary coverage and scenario diversity. Collectively, these resources have greatly accelerated the progress of SLR, establishing shared benchmarks for model comparison and advancing both word-level and sentence-level SLR understanding.

Despite these achievements, vision-only approaches face fundamental limitations that restrict their scalability and real-world deployment: 
(1) They are highly sensitive to environmental conditions; the accuracy deteriorates under poor lighting, cluttered backgrounds, or self-occlusion.
(2) They raise privacy concerns, as video data inevitably expose identifiable facial and bodily information, posing ethical challenges in settings such as classrooms, clinics, and homes.
(3) They suffer from limited deployment flexibility, since cameras typically require fixed viewpoints and unobstructed fields of view, making them unsuitable for pervasive or mobile scenarios.
These challenges highlight the need for multimodal sensing that can complement visual input with, for examples, depth, radar, or inertial signals, paving the way toward more robust, privacy-preserving, and ubiquitous sign language recognition systems.

To overcome these limitations, researchers have increasingly explored non-visual sensing modalities for sign language recognition, e.g., millimeter-wave (mmWave) radar and inertial measurement units (IMUs).
mmWave radar captures body dynamics even under occlusion or low-light conditions~\cite{santhalingam2020expressive,lan2023applying,gurbuz2020american}, while avoiding identifiable visual cues, thereby offering inherent privacy advantages. Its fine-grained motion sensitivity also makes it well-suited for detecting subtle hand gestures and finger movements.
IMUs, on the other hand, are wearable motion sensors that record acceleration and angular velocity at high frequencies~\cite{zhang2022wearsign,santhalingam2023synthetic,hou2019signspeaker,jin2023smartasl,khomami2021persian}, providing first-person kinematic data that directly reflect signer movements. IMUs are low-cost, lightweight, and privacy-preserving, making them particularly suitable for continuous and unobtrusive daily use. The complementarity properties across these modalities provide strong motivation for multimodal fusion and cross-modality learning.
Although multimodal approaches have shown effectiveness in related domains such as human activity recognition (HAR) \cite{zhu2025master,shi2025human,gao2023real,guo2016wearable,kong2019mmact}, the sign language recognition community still lacks large-scale, synchronized datasets that integrate vision, radar, and inertial sensing.

\begin{table}[h]
\caption{
Comparison between \systemname{} and existing isolated sign language recognition (ISLR) datasets or methods. 
Entries marked with \textbf{(D)} denote publicly released datasets. 
Language acronyms are listed in Appendix~\ref{app:language_acronyms}.
}
\label{tab:ISLR datasets}
\centering
\begin{tabularx}{\linewidth}{c c *{4}{c} r r c}
\toprule
\multirow{2}{*}{\textbf{Dataset (D)/Methods}} &
\multirow{2}{*}{\textbf{Language}} &
\multicolumn{4}{c}{\textbf{Modalities}} &
\multirow{2}{*}{\textbf{Signs}} &
\multirow{2}{*}{\textbf{Signers}} &
\multirow{2}{*}{\textbf{Source}} \\
\cmidrule(lr){3-6}
& & RGB & Depth & IMU & mmWave & & & \\
\midrule
Purdue RVL-SLLL~\cite{martinez2002purdue}(\textbf{D})  & ASL & \cmark & \xmark & \xmark & \xmark & 39 & 14 & Studio \\
RWTH-BOSTON 50~\cite{zahedi2005combination}(\textbf{D})  & ASL & \cmark & \xmark & \xmark & \xmark & 50 & 3 & Studio \\
ASLLVD~\cite{athitsos2008american}(\textbf{D})  & ASL & \cmark & \xmark & \xmark & \xmark & 3000 & 6 & Studio \\
WLASL~\cite{li2020word}(\textbf{D}) & ASL & \cmark & \xmark & \xmark & \xmark & 2000 & 119 & Web \\
MS-ASL~\cite{joze2018ms}(\textbf{D})  & ASL & \cmark & \xmark & \xmark & \xmark & 1000 & 222 & Web \\
ASL Citizen~\cite{desai2023asl}(\textbf{D})  & ASL & \cmark & \xmark & \xmark & \xmark & 2731 & 52 & Webcam \\
PopSign ASL v1.0~\cite{starner2023popsign}(\textbf{D})  & ASL & \cmark & \xmark & \xmark & \xmark & 250 & 47 & Smartphone \\
BSL-1K~\cite{albanie2020bsl}(\textbf{D})  & BSL & \cmark & \xmark & \xmark & \xmark & 1064 & 40 & Web \\
DEVISIGN-L~\cite{chai2013sign}(\textbf{D})  & CSL & \cmark & \cmark  & \xmark & \xmark & 2000 & 8 & Studio \\
CSL 500~\cite{huang2018attention}(\textbf{D}) & CSL & \cmark & \cmark & \xmark & \xmark & 500 & 50 & Studio \\
SMILE~\cite{ebling2018smile}(\textbf{D})  & DSGS & \cmark & \cmark & \xmark & \xmark & 100 & 30 & Studio \\
GSL 982~\cite{ong2012sign}(\textbf{D})  & GSL & \cmark & \cmark & \xmark & \xmark & 982 & 1 & Studio \\
INCLUDE~\cite{sridhar2020include}(\textbf{D})  & ISL & \cmark & \xmark & \xmark & \xmark & 263 & 7 & Studio \\
LSA 64~\cite{ronchetti2023lsa64}(\textbf{D})  & LSR & \cmark & \xmark & \xmark & \xmark & 64 & 10 & Studio \\
LSE-Sign~\cite{gutierrez2016lse}(\textbf{D})  & LSE & \cmark & \xmark & \xmark & \xmark & 2400 & 2 & Studio \\
LSFB-ISOL~\cite{fink2021lsfb}(\textbf{D})  & LSFB & \cmark & \xmark & \xmark & \xmark & 395 & 100 & Studio \\
BosphorusSign22K~\cite{ozdemir2020bosphorussign22k}(\textbf{D})  & TSL & \cmark & \cmark & \xmark & \xmark & 744 & 6 & Studio \\
AUTSL~\cite{sincan2020autsl}(\textbf{D})  & TSL & \cmark & \cmark & \xmark & \xmark & 226 & 43 & Studio \\
MM-WLAusian~\cite{shen2024mm}(\textbf{D})  & Auslan & \cmark & \cmark & \xmark & \xmark & 3215 & 73 & Studio \\ \midrule
PSL-IMU-sEMG~\cite{khomami2021persian}  & PSL & \xmark & \xmark & \cmark & \xmark & 20 & 10 & Studio \\
SignSpeaker~\cite{hou2019signspeaker}  & ASL & \xmark & \xmark & \cmark & \xmark & 129 & 16 & Studio \\ \midrule
ExASL~\cite{santhalingam2020expressive}  & ASL & \xmark & \cmark & \xmark & \cmark & 23 & 5 & Studio \\
RadarSign~\cite{lan2023applying}  & CSL & \xmark & \xmark & \xmark & \cmark & 15 & 3 & Studio \\
Rahman et al. (2021)~\cite{rahman2021word}  & ASL & \xmark & \xmark & \xmark & \cmark & 20 & 15 & Studio \\
RF-sign~\cite{gurbuz2020american}  & ASL & \xmark & \xmark & \xmark & \cmark & 20 & 13 & Studio \\
Gurbuz et al. (2020)~\cite{gurbuz2020linguistic}  & ASL & \cmark & \cmark & \xmark & \cmark & 20 & 3 & Studio \\
mmASL~\cite{santhalingam2020mmasl}(\textbf{D})  & ASL & \cmark & \cmark & \xmark & \cmark & 50 & 15 & Studio \\ \midrule
\systemname{}(\textbf{D}) & \textbf{ASL} & \textbf{\cmark} & \textbf{\cmark} & \textbf{\cmark} & \textbf{\cmark} & \textbf{160} & \textbf{20} & \textbf{Studio} \\
\bottomrule
\end{tabularx}
\end{table}

To underscore the necessity of large-scale multimodal datasets for advancing SLR research, we conducted an extensive literature review of recent SLR studies and summarized the representative datasets and methods in Table~\ref{tab:ISLR datasets}. Works that publicly released their collected datasets are marked with $(\mathbf{D})$. The table provides key information for each dataset or method across multiple dimensions. Specifically, the second column lists the language used in each study, including American Sign Language (ASL), British Sign Language (BSL), Chinese Sign Language (CSL), and others (the full list of the types of sign languages can be found in appendix \ref{app:language_acronyms}). The middle columns indicate the modalities employed, with check marks denoting the inclusion of each sensing type, followed by the number of signs and number of signers. The last column specifies the data collection source, distinguishing between datasets captured in controlled environments (studio), collected from online resources (Web), or recorded via consumer-grade devices such as webcams or smartphones.

From Table~\ref{tab:ISLR datasets}, we observe that most existing ISLR datasets rely exclusively on vision-based modalities such as RGB or RGB-D cameras, with limited exploration of non-visual sensing. A few recent studies have introduced mmWave radar or IMU sensors; however, these datasets remain small in scale, restricted in vocabulary, and often unavailable to the public. The only notable exception is the mmASL dataset, which combines visual and mmWave modalities but covers only a limited number of signs. Moreover, none of the existing resources achieve comprehensive multimodal integration across RGB, depth, radar, and inertial signals. This imbalance reveals a critical research gap: while large-scale visual datasets have driven remarkable progress in model accuracy, advances in multimodal fusion, cross-sensor correlation, and privacy-preserving sign recognition have been constrained by the absence of large-scale, high-quality multimodal benchmarks.

To bridge this gap, we present \systemname{}, a new multimodal dataset for SLR, as shown in the last row of Table~\ref{tab:ISLR datasets}. The dataset was collected in a controlled studio environment, where participants were seated in front of an Azure Kinect RGB-D camera and an mmWave radar, while wearing IMUs on both wrists. All sensing devices were connected to a single workstation to achieve millisecond-level synchronization across modalities. This setup enables complementary sensing, capturing semantically rich RGB-D frames from the Kinect, fine-grained hand and arm motion reflections from the mmWave radar, and precise kinematic measurements from the IMUs. To assess its utility, we further conduct comprehensive benchmarking experiments using state-of-the-art recognition models under both single-modality and multimodal fusion settings.

\noindent The contributions of this work are as follows:

\begin{itemize}
    \item {\bf A large-scale multimodal sign language dataset:} {\systemname{}}\footnote{The dataset is publicly available at \url{https://github.com/happy2sumture-cloud/SIGMA-ASL}} is a synchronized multimodal dataset for SLR, integrating RGB-D, mmWave radar, and wearable IMU sensing. The dataset includes recordings from 20 participants performing 160 commonly used ASL signs, resulting in over 93,000 word-level clips, making it one of the most comprehensive multimodal resources for sign language research to date. The dataset will be released to the public upon acceptance. 
    \item {\bf A unified collection and synchronization framework:} We develop a high-precision data collection framework that achieves millisecond-level synchronization across heterogeneous sensors, ensuring temporal and spatial alignment between visual, radar, and inertial modalities. This setup enables reliable multimodal fusion and cross-modal learning research.
    \item {\bf Comprehensive preprocessing and benchmarking:} We design standardized preprocessing pipelines and establish benchmarking protocols under both user-dependent and user-independent settings. Extensive evaluations are conducted on state-of-the-art models across single and fused modalities, providing a systematic performance baseline for future multimodal SLR research.
\end{itemize}

\section{Related Work}

This section reviews the most relevant studies to \systemname{}. Since \systemname{} primarily provides a multimodal dataset of word-level sign clips, our review focuses on three major areas: (1) isolated sign language recognition (ISLR) datasets, (2) ISLR methodologies across different sensing modalities, and (3) multimodal action and gesture recognition.

\subsection{ISLR Datasets}

Over the past decades, numerous datasets have been developed to advance ISLR research; however, most remain limited in modality diversity, spatial coverage, and vocabulary richness, as summarized in Table~\ref{tab:ISLR datasets}.
Early benchmark datasets such as Purdue RVL-SLLL~\cite{martinez2002purdue} and ASLLVD~\cite{athitsos2008american} laid the groundwork for ASL research but primarily relied on single-view RGB recordings without depth or motion cues, restricting the modeling of three-dimensional signing dynamics. Later large-scale datasets, including WLASL~\cite{li2020word} and MS-ASL~\cite{joze2018ms}, expanded the number of glosses and signers but still relied solely on 2D visual data, making them susceptible to occlusions and viewpoint variations. In contrast, datasets such as CSL 500~\cite{huang2018attention} and GSL 982~\cite{ong2012sign} introduced depth sensing to better represent spatial structure; however, their limited vocabulary restricts the usability.
More recently, multimodal ISLR datasets have been introduced to bridge the gap between visual perception and motion sensing. Nevertheless, existing multimodal datasets remain small in scale, constrained in modality combinations, and often lack precise temporal synchronization. The mmASL dataset~\cite{santhalingam2020mmasl} represents a notable example by combining visual and mmWave modalities, though its vocabulary size is limited. 

\subsection{Isolated Sign Language Recognition Methods}
\subsubsection{Vision-based ISLR}

Traditional vision-based ISLR methods primarily rely on convolutional and recurrent neural networks to learn spatiotemporal representations from RGB videos. Early works employed architectures such as C3D~\cite{tran2015learning}, I3D~\cite{carreira2017quo}, and 3D-ResNet~\cite{hara2018can}. Later approaches leveraged graph convolutional networks (GCNs), including ST-GCN~\cite{yan2018spatial} and PoseC3D~\cite{duan2022revisiting}, to explicitly model human skeletal topology and temporal dependencies among body keypoints. More recently, Transformer-based architectures~\cite{dosovitskiy2020image,bertasius2021space,sui2024tramba} have shown superior capabilities in long-range temporal modeling and contextual reasoning, improving recognition robustness across signer variations and temporal dynamics.
Despite these advances, purely vision-based systems remain highly sensitive to illumination changes, background clutter, and occlusion.

\subsubsection{IMU-based ISLR}

IMUs offer a lightweight, privacy-preserving means of capturing motion dynamics through tri-axial acceleration and angular velocity. Early methods extracted handcrafted statistical features and employed classical classifiers for recognition \cite{yuan2020hand,nguyen2021gesture}. With the rise of deep learning, models such as recurrent neural networks (RNNs)~\cite{zaremba2014recurrent}, temporal convolutional networks (TCNs)~\cite{bai2018empirical}, and attention-based architectures~\cite{singh2020deep} have become dominant, enabling end-to-end learning directly from raw IMU sequences.
IMU-based approaches are inherently robust to lighting variation and occlusion, making them suitable for wearable and mobile contexts. However, their lack of explicit spatial information makes it difficult to disambiguate signs with similar trajectories but different handshapes or orientations.

\subsubsection{mmWave-based ISLR}

mmWave radar captures fine-grained motion information by analyzing range–Doppler or range–angle–Doppler signatures of reflected electromagnetic signals, maintaining performance even under non-line-of-sight (NLOS) conditions. Recent works, including RadarSign~\cite{lan2023applying} and RF-sign~\cite{gurbuz2020american}, have demonstrated the feasibility of isolated sign or gesture recognition using mmWave radar, typically leveraging CNN- or RNN-based encoders to extract micro-Doppler temporal features. 

\subsection{Multi-modal Action and Gesture Recognition}
Although multimodal learning has been less explored in SLR, it has been extensively investigated in the domains of human action and gesture recognition, where researchers integrate complementary cues from diverse sensing modalities such as RGB videos, depth maps, IMUs, audio, and mmWave radar.

Large-scale multimodal datasets have significantly advanced research on HAR. The NTU RGB+D dataset~\cite{shahroudy2016ntu} includes over 56,000 RGB-D video samples with accurate 3D skeleton annotations, highlighting the benefits of integrating geometric and visual information. PKU-MMD~\cite{liu2017pku} extended this paradigm to continuous, multi-view, and multi-modality action recognition by combining RGB, depth, infrared, and skeleton data under consistent calibration. MMAct~\cite{kong2019mmact} further bridged visual and wearable sensing modalities by integrating acceleration, gyroscope, orientation, Wi-Fi, and pressure signals, providing a foundation for cross-modal activity understanding.

To effectively integrate heterogeneous sensing inputs, a variety of multimodal fusion paradigms have been explored. Early fusion \cite{team2024chameleon,zhang2024evf} combines features from multiple modalities at the input or embedding level to enable end-to-end joint optimization. Late fusion \cite{feichtenhofer2016convolutional,liu2018late} aggregates outputs from modality-specific models through weighted averaging, voting, or attention mechanisms, offering robustness to missing or unreliable data. Hybrid and intermediate fusion strategies \cite{tsai2019multimodal}, often implemented using transformer-based architectures, further enable fine-grained temporal alignment and adaptive cross-modal weighting.
Despite these advances, sign language recognition still lacks large-scale, time-synchronized multimodal datasets. Existing corpora such as WLASL~\cite{li2020word}, MS-ASL~\cite{joze2018ms}, and CSL 500~\cite{huang2018attention} remain predominantly visual, while smaller datasets incorporating IMU or EMG signals are limited in both vocabulary size and data diversity. However, to date, no publicly available resource integrates RGB, depth, IMU, and mmWave radar within a unified and precisely synchronized framework like \systemname{}.

\section{The \systemname{} Dataset}

In this section, we first describe the selection protocol of the sign language vocabulary and the setup of our synchronized tri-sensor recording platform, which consists of a Kinect RGB-D camera, an mmWave radar, and two wearable IMUs. We then detail the data collection workflow, including data preprocessing, sensor synchronization. At last the important statistics of the dataset are presented.

\subsection{Vocabulary Selection}

Existing multimodal SLR datasets remain highly limited in both scale and vocabulary diversity; for example, the recent mmASL dataset~\cite{santhalingam2020mmasl} includes only 50 signs. This scarcity primarily arises from the high cost and logistical complexity of multimodal data collection, which demands specialized sensing hardware, precise sensor calibration, and time-consuming recording procedures for each participant. Consequently, vocabulary selection in multimodal datasets must carefully balance linguistic coverage with practical feasibility, while ensuring that the chosen signs exhibit sufficient kinematic distinctiveness to enable reliable recognition.

\begin{figure}[htbp]
    \centering
        \begin{subfigure}{0.45\linewidth}
        \centering
        \includegraphics[width=\linewidth]{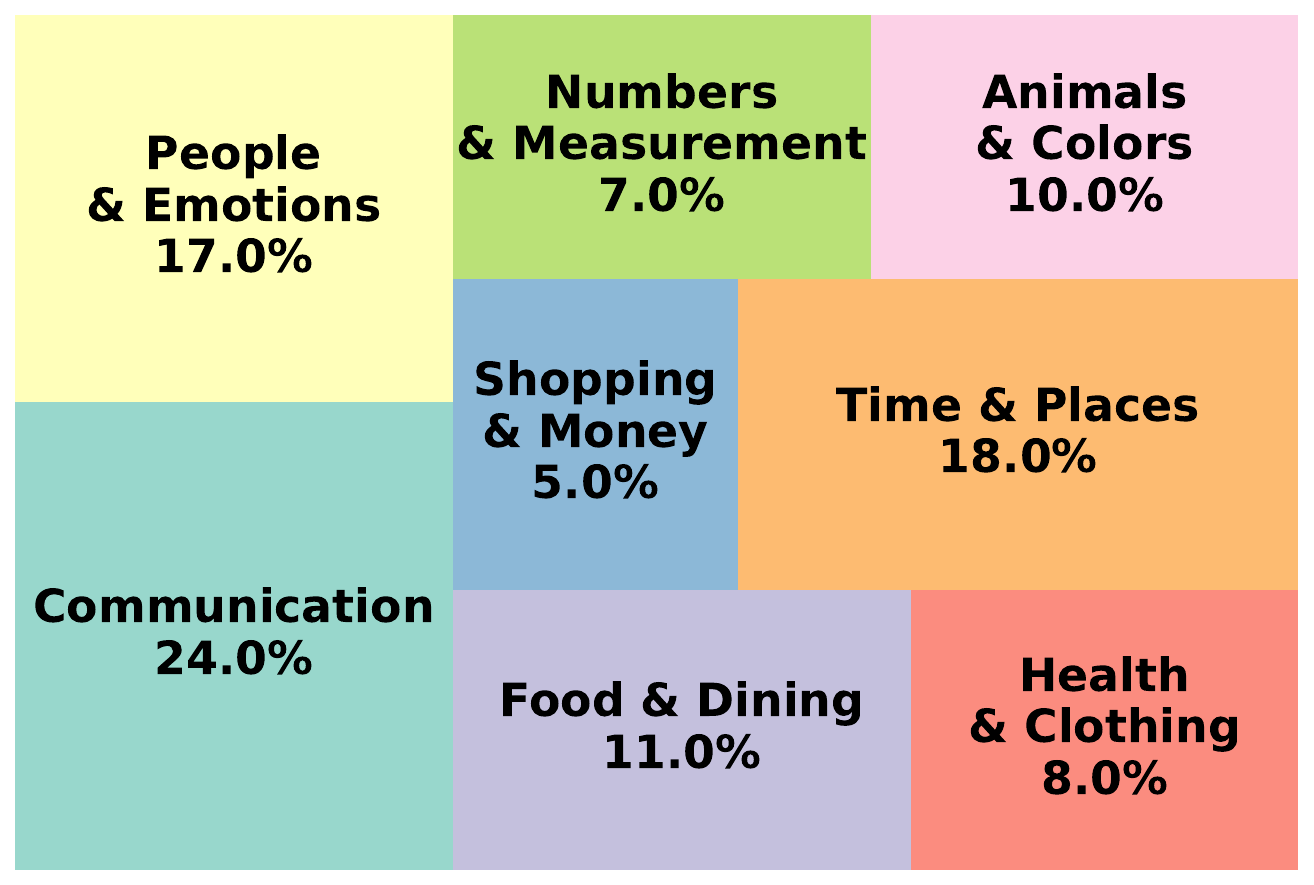}
        \caption{Semantic Categories.}
        \label{fig:category_distribution}
    \end{subfigure}
    \begin{subfigure}{0.35\linewidth}
        \centering
        \includegraphics[width=\linewidth]{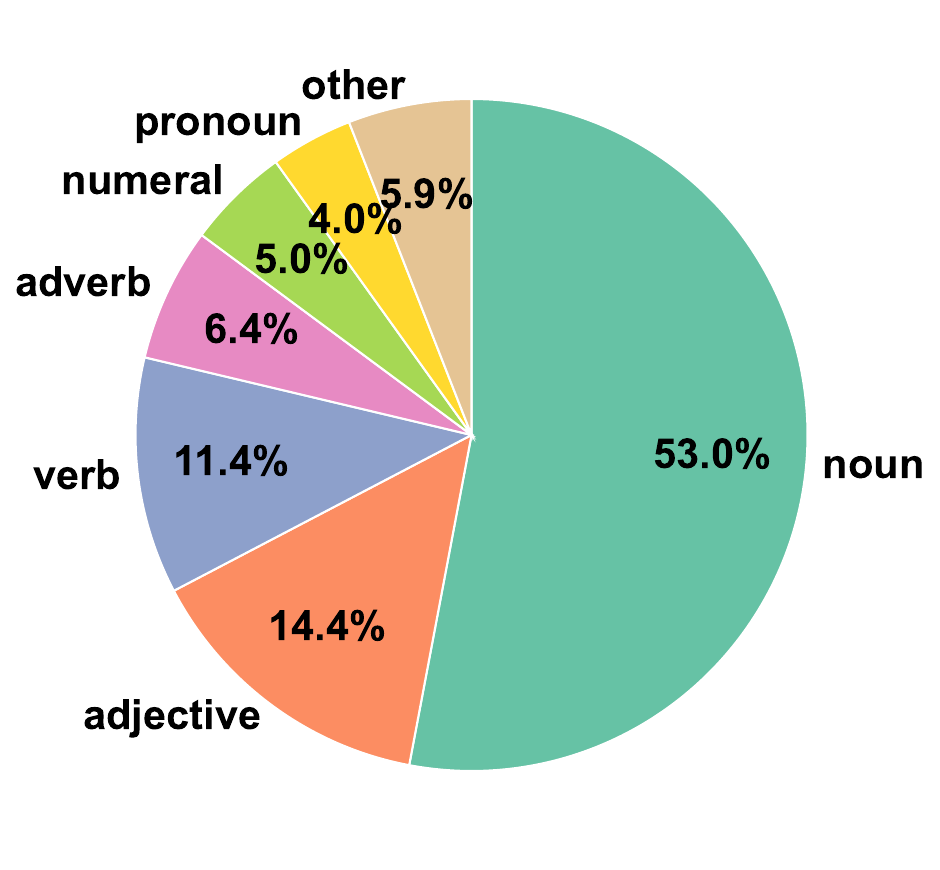}
        \caption{Part-of-Speech Frequency.}
        \label{fig:pos_counts_bar}
    \end{subfigure}

    \caption{The distributions of semantic categories (a) and POS (b) of the selected vocabulary.}
    \label{fig:pos_category_combined}
\end{figure}

To address the above considerations, we adopted a systematic vocabulary selection strategy for the \systemname{} dataset. We prioritized pedagogical relevance and frequency of use by incorporating the 103 most commonly taught ASL words from the ASL University curriculum \cite{vicars_first100signs} .\added{~ Recognizing that a single ASL gloss may admit multiple gesture realizations, we additionally included alternative forms for 21 of these words, yielding 21 semantically equivalent but kinematically distinct variants.} Although these variants convey similar meanings, they \added{differ in handshape, articulation, and/or motion dynamics; thus, we treat each form as an independent sign instance to facilitate} a \added{rigorous} evaluation of recognition robustness and intra-lexical variability across modalities. \added{We further included 26 alphabet letters and 10 digits as foundational components of everyday ASL communication: fingerspelling enables the production of proper nouns (e.g., names and place names) and out-of-vocabulary terms, while number signs frequently occur in daily interactions (e.g., age, quantity, and time expressions), aligning with the same high-frequency, practical-usage rationale as the selected common words.} The final vocabulary comprises 160 signs spanning multiple semantic domains and parts of speech (POS) such as nouns, verbs, and adverbs\added{~(see Appendix~\ref{app:vocabulary_list} for the full list)}. Overall, the dataset contains 70 two-handed and 90 one-handed signs, ensuring both linguistic diversity and kinematic variety suitable for multimodal sensor-based recognition.

As shown in Fig.~\ref{fig:pos_category_combined}, the vocabulary distributions across semantic categories and POS frequencies demonstrate broad linguistic and conceptual coverage within the \systemname{} dataset. The selected words span diverse everyday contexts, such as communication, people and emotions, time and places, and food, capturing both static and dynamic signing patterns. Nouns constitute the majority (53\%), followed by adjectives, verbs, and adverbs, reflecting the natural composition of frequently used ASL lexicons.

\subsection{Data Collection Setup and Procedure}

\added{
We recruited 20 participants (7 female, 13 male; ages 20–30) for data collection.\footnote{All experimental procedures were approved by the local IRB council.} Participants reviewed an electronic informed-consent form and provided signed consent via print or electronic signature. Each participant received approximately USD 50 in compensation. A de-identified consent template is included in the supplementary materials. All participants were hearing novice ASL learners with no prior formal sign-language training, reflecting realistic assistive-technology deployment scenarios; consequently, the recordings primarily capture pedagogical or imitated signing rather than naturally occurring signing within Deaf or hard-of-hearing (HoH) communities.
} Prior to recording, each participant completed standardized training sessions in which they watched reference videos \cite{vicars_first100signs} and practiced the target signs under the supervision of the experimenter.

\begin{figure}[htb]
\centering
\includegraphics[width=0.7\linewidth]{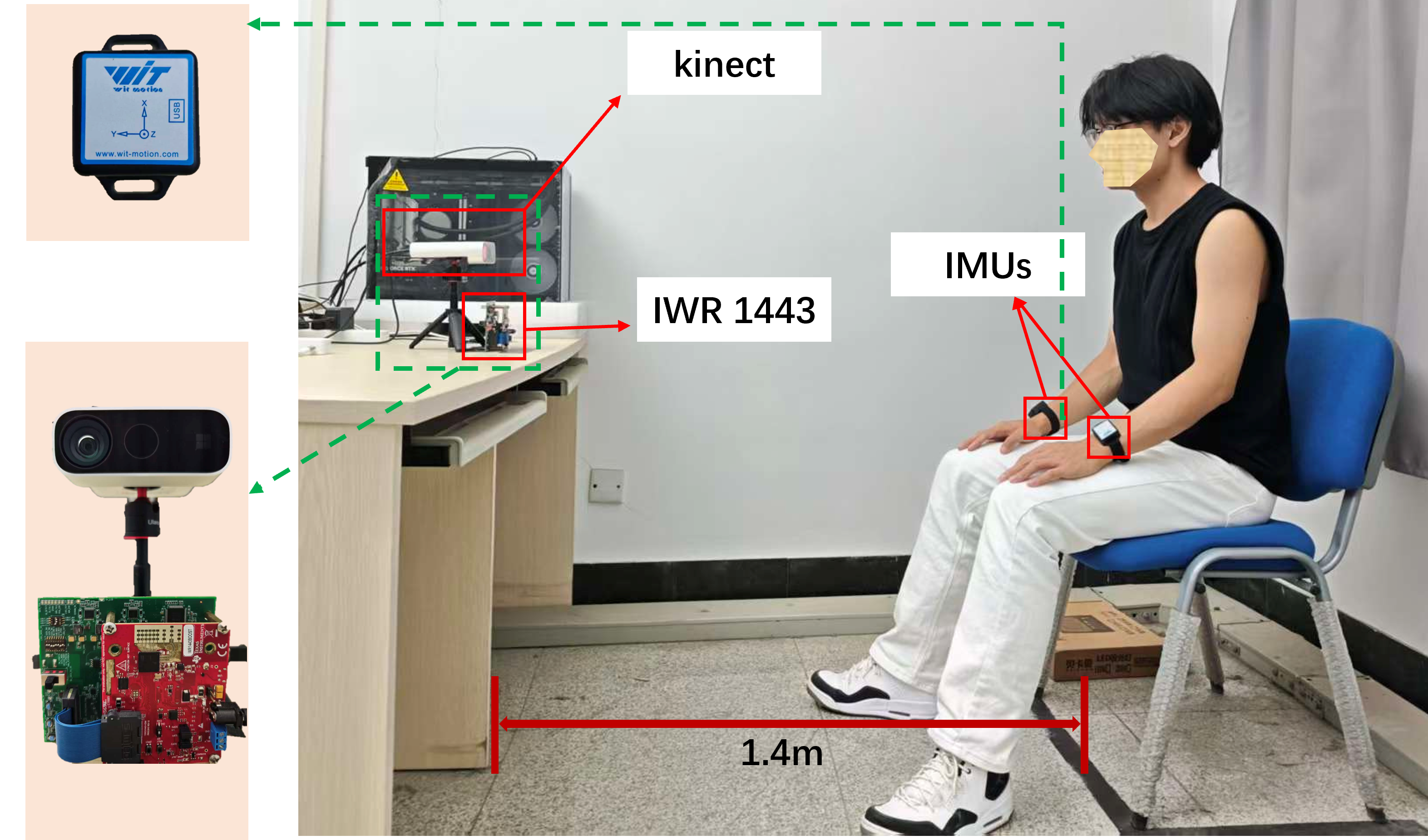}
\caption{Data collection setup and sensor deployment.}
\label{fig:collection_setup}
\end{figure}

\begin{table}[htbp]
\centering
\caption{
Comparison of different sensors used in the multimodal system. 
\textit{Intrusive} specifies whether the sensor requires physical contact with the user. 
Prices are approximate as of 2025 (in USD).
}
\label{tab:setup_comparison}
\begin{tabular}{@{}lccccc@{}}
\toprule
\textbf{Sensor} & \textbf{Model} & \textbf{Sampling Rate (Hz)} & \textbf{Intrusive} & \textbf{Price} \\ 
\midrule
mmWave  & TI IWR1443~\cite{TI_IWR1443BOOST}  & 30 & No  & \$358.81 \\
Kinect  & Azure Kinect~\cite{AzureKinectDK}  & 30 & Yes & \$399 \\
IMU     & WT901~\cite{witmotion_bwt901cl}  & 100 & No  & \$7 ea. \\
\bottomrule
\end{tabular}
\end{table}

To collect the multimodal dataset, we developed a synchronized sensing system consisting of an Azure Kinect RGB-D camera, a millimeter-wave (mmWave) radar, and two wrist-worn IMU sensors 
. Detailed specifications of each device are provided in Table~\ref{tab:setup_comparison}. 
To achieve precise temporal alignment across modalities, we adopted a unified synchronization framework inspired by~\cite{an2022mri}. 
All sensors were connected to a single host computer that provided a unified system clock for timestamping. Each data stream was recorded with the host’s local time reference to ensure consistent cross-sensor synchronization. As shown in Fig.~\ref{fig:collection_setup}, the experimental setup included multiple synchronized sensors positioned to capture comprehensive multimodal motion data. The Kinect RGB-D camera was mounted on a table approximately 1.4 meters from the participant at torso height, providing a frontal view for tracking upper-body and hand movements. The mmWave radar was placed adjacent to the Kinect to maintain a similar field of view and spatial coverage. To capture detailed wrist-level motion, two WT901 IMUs were fastened to the participant’s left and right wrists using adjustable straps.
The sensing system employed heterogeneous communication protocols optimized for each modality: the Kinect and mmWave radar used wired USB connections to ensure high-bandwidth data transfer with minimal latency, while the IMU sensors communicated wirelessly via Wi-Fi, allowing participants' hands to move freely without restricting natural signing gestures.

During data collection, each participant completed two recording sessions encompassing all 160 ASL vocabulary words, scheduled on separate days to mitigate fatigue effects and capture natural temporal variations in signing behavior, hand positioning, and sensor attachment. This cross-session protocol enables the evaluation of model robustness under realistic conditions, where day-to-day variability can significantly influence recognition performance. For each target word, participants were allocated a 30-second recording window to repeatedly perform the corresponding sign. To facilitate automated segmentation, participants were instructed to assume a neutral rest pose, i.e., hands resting on thighs, at the beginning and end of each trial, providing clear temporal boundary markers. The system required a minimum of 12 repetitions per word; if the number of valid samples fell below this threshold, supplementary recordings were immediately conducted to ensure at least 10 synchronized samples across all modalities (the samples at the beginning and end might be dropped due to temporal misalignment).
During acquisition, participants sat comfortably with IMU sensors securely fastened to both wrists. Upon system initialization, the experimenter issued verbal ``Start'' and ``Stop'' cues to control the recording window. Real-time quality monitoring was performed throughout data collection, and any sessions not meeting the repetition or synchronization requirements were promptly re-recorded before proceeding to the next item. All recordings were conducted in a controlled indoor environment with consistent lighting and temperature conditions to ensure participant comfort and minimize environmental variability across sessions.

\subsection{Data Preprocessing}

After collecting the raw multimodal spatio-temporal recordings from all participants, a series of preprocessing procedures were applied to prepare the data for subsequent analysis and evaluation. Specifically, the recordings were segmented into individual word clips, spectrograms were generated from the mmWave radar signals and the IMU time-series data were transformed into the frequency domain to better capture motion dynamics and periodic patterns. Finally, outlier samples resulting from synchronization errors or transmission failures were identified and removed to ensure high data quality and consistency across all modalities.

\subsubsection{Multimodal Word Clips Segmentation.}
As described in the data collection procedure above, each multimodal 30-second recording contains multiple repetitions of the target sign, separated by brief neutral rest poses. To facilitate convenient use for future research, these recordings were segmented into individual word-level clips. Rather than relying on time-consuming manual annotation, we developed an automated segmentation approach that leverages RGB imagery to detect hand motion patterns. The extracted hand-activity sequences were then converted into binary temporal signals, i.e., either in moving (``1'') or resting (``0'') states, which were analyzed to identify sign boundaries efficiently and consistently across all recordings.

\begin{figure}[h]
    \centering
    \begin{subfigure}{0.28\linewidth}
        \includegraphics[width=\linewidth]{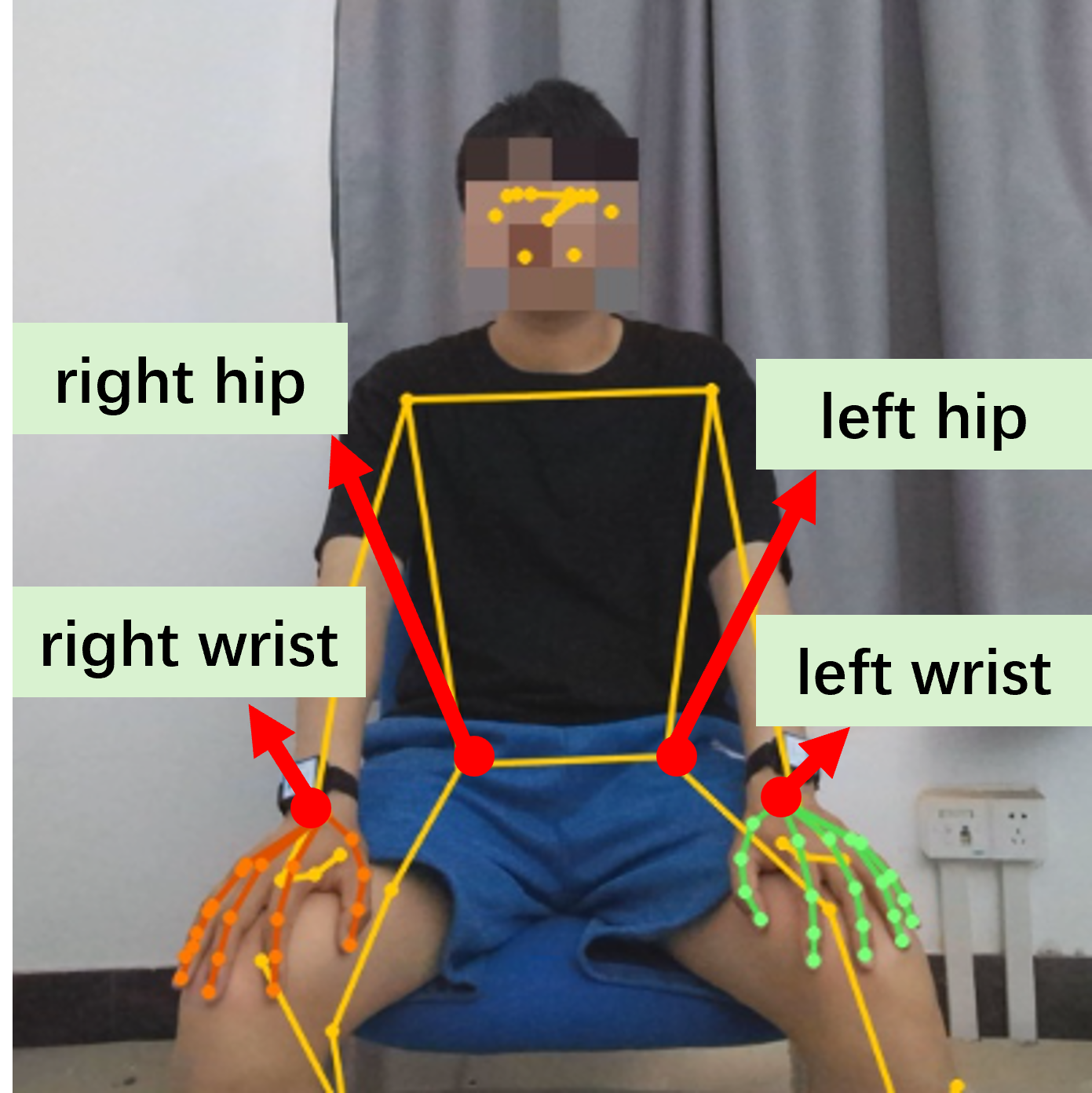}
        \caption{\centering
        Landmarks extracted \\from MediaPipe}
        \label{fig:body_hands_landmarks}
    \end{subfigure}
     \begin{subfigure}{0.7\linewidth}
        \includegraphics[width=\linewidth]{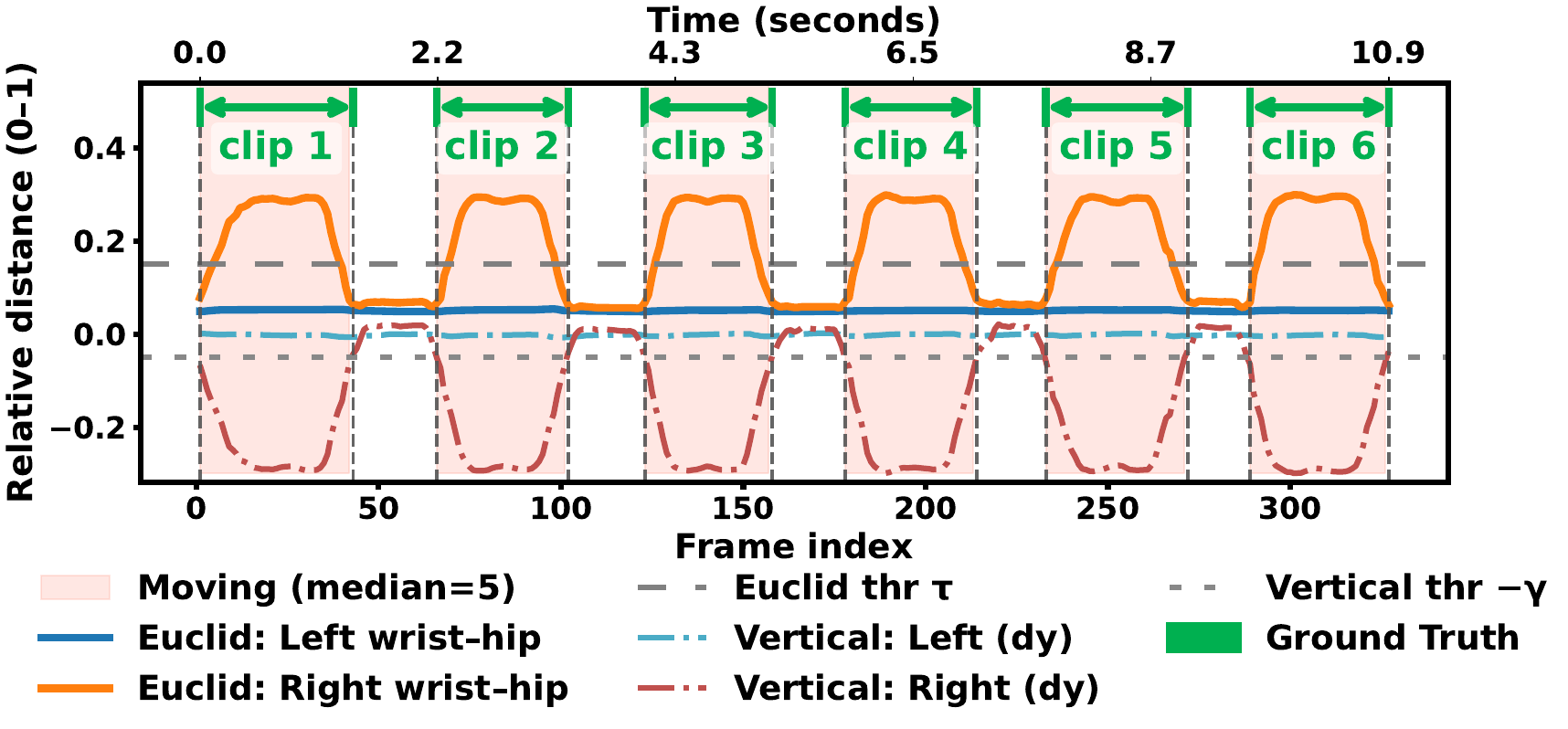}
        \caption{\centering
        Wrist–hip keypoint Euclidean distance and vertical offset\\ used for clip boundary detection.}
        \label{fig:keypoint_trajectory}
    \end{subfigure}
    \caption{Landmarks are extracted from RGB frames and the offsets between the key points (wrist-hip) are used to find the boundaries between the word clips.}
    \label{fig:placeholder}
\end{figure}

Specifically, for each 30-second recording session, we extracted 33 full-body keypoints from the RGB frames using MediaPipe Pose~\cite{lugaresi2019mediapipe}, which provided skeletal landmarks including shoulders, elbows, wrists, hips, and other anatomical points. As the sign languages are mostly represented as hand gestures, we extracted 21 keypoints per hand using MediaPipe Hands~\cite{zhang2020mediapipe}, capturing detailed finger joint positions and palm landmarks for both left and right hands as shown in Fig. \ref{fig:body_hands_landmarks}.

To ensure temporal consistency and address occlusions or detection failures, we implemented a post-processing strategy to refine the keypoint outputs from MediaPipe. When MediaPipe failed to detect keypoints in a given frame, typically due to motion blur, occlusion, or complex hand poses, the missing values were replaced with zero placeholders to explicitly mark undetected points. To preserve temporal smoothness, the most recent valid keypoint coordinates were propagated forward until new detections became available, effectively bridging short-term detection gaps and maintaining stable motion trajectories across frames.

By tracking the detected keypoints across consecutive frames, we obtained the temporal trajectories of hand and body motion. To automatically determine the boundaries of word-level clips, we computed two distance-based features that represent the relative position of each hand with respect to the body. As illustrated in Fig.~\ref{fig:keypoint_trajectory}, we calculated both the 2D Euclidean distance and the vertical offset between each wrist and its corresponding hip (i.e., left wrist–left hip and right wrist–right hip). A frame was classified as a static state only when both wrists simultaneously satisfied two criteria: the wrist does not rise significantly above the hip (the vertical difference remains within a predefined threshold), and the wrist stays sufficiently close to the hip (the spatial distance falls below a specified threshold). Otherwise, the frame was labeled as a moving state, indicating that the hands are engaged in signing. The transitions between static and moving states provide reliable temporal cues for locating the start and end of individual sign words, as shown in Fig.~\ref{fig:keypoint_trajectory}. To mitigate short-term fluctuations, a median filter with a window length of five frames was applied to the binary state sequence. Using this procedure, we obtained a total of \added{93,545} word-level clips across all sensing modalities, which were subsequently utilized for downstream processing and evaluation.

\subsubsection{Range-Doppler Map (RDM) Calculation for mmWave Radar Signals }

\begin{figure}[h]
\centering
\begin{subfigure}[t]{0.48\linewidth}
    \centering
    \small                                 %
    \setlength{\tabcolsep}{4pt}            %
    \begin{tabular}{@{}llc@{}}
        \toprule
        \textbf{Parameter} & \textbf{Description} & \textbf{Value} \\ 
        \midrule
        $N_{\mathrm{tx}}$   & Number of transmit antennas          & 3 \\
        $N_{\mathrm{rx}}$   & Number of receive antennas           & 4 \\
        $f_{\mathrm{start}}$& Start frequency (GHz)                & 77 \\
        $S_{\mathrm{freq}}$ & Frequency slope (MHz/$\mu$s)         & 61.508 \\
        $N_{\mathrm{ADC}}$  & ADC samples per chirp                & 256 \\
        $f_{\mathrm{s}}$    & Sampling rate (ksps)                 & 5000 \\
        $L_{\mathrm{frame}}$& Loops per frame                      & 128 \\
        $T_{\mathrm{period}}$& Frame periodicity (ms)              & 33.33 \\
        \bottomrule
    \end{tabular}
    \caption{Radar configuration parameters.}
    \label{fig:radar_params}
\end{subfigure}
\hfill
\begin{subfigure}[t]{0.48\linewidth}
    \centering
    \raisebox{-0.7in}{\includegraphics[width=0.95\linewidth]{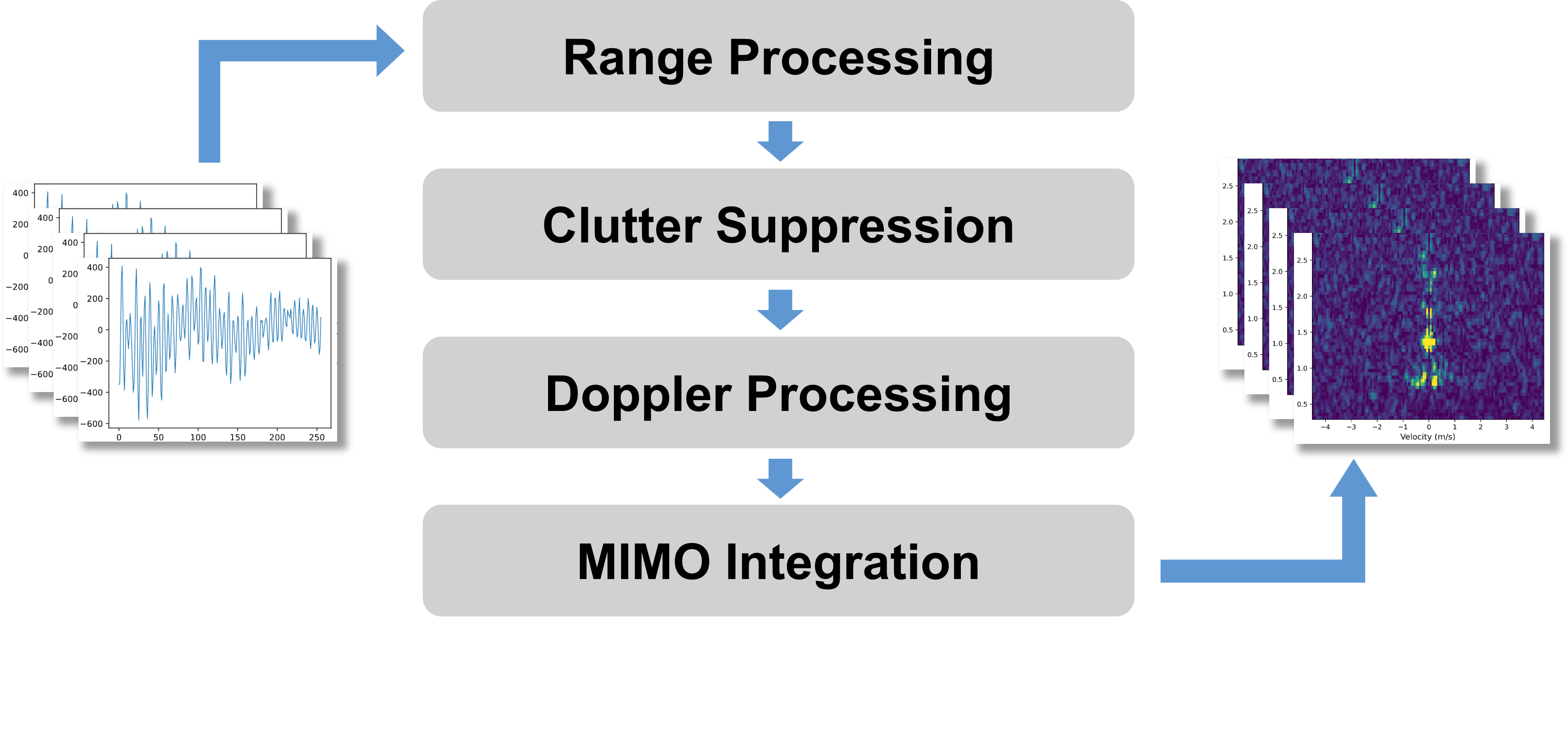}}
    \caption{Radar signal pre-processing pipeline. }
    \label{fig:radar_pipeline}
\end{subfigure}

\caption{Radar module overview: (a) hardware configuration parameters: (TI IWR1443 FMCW mmWave) and (b) the pre-processing pipeline: Range Processing, Clutter Suppression, Doppler Processing and MIMO Integration to form the RDM.}
\label{fig:radar_table_and_pipeline}
\end{figure}

 During data collection, we employed the Texas Instruments IWR1443BOOST evaluation module~\cite{TI_IWR1443BOOST}, 
which integrates a 77~GHz Frequency-Modulated Continuous Wave (FMCW) radar frontend with on-chip signal processing capabilities. The detailed configuration parameters used in our experiment are summarized in the left of Fig.~\ref{fig:radar_table_and_pipeline}. Following demodulation and de-interleaving, the captured radar waveforms were reconstructed into a five-dimensional complex-valued data cube $\mathbf{X} \in \mathbb{C}^{S \times N_{\mathrm{rx}} \times N_{\mathrm{tx}} \times L \times T}$, where $S{=}256$ denotes the number of fast-time ADC samples per chirp, $N_{\mathrm{rx}}{=}4$ and $N_{\mathrm{tx}}{=}3$ represent the number of receive and transmit antennas, respectively, $L{=}128$ corresponds to the number of chirps per frame, and $T$ indexes the consecutive temporal frames.

During data collection, we employed the Texas Instruments IWR1443BOOST evaluation module~\cite{TI_IWR1443BOOST},
which integrates a 77~GHz Frequency-Modulated Continuous Wave (FMCW) radar frontend with on-chip signal processing capabilities. The detailed configuration parameters used in our experiment are summarized in the left of Fig.~\ref{fig:radar_table_and_pipeline}. Following demodulation and de-interleaving, the captured radar waveforms were reconstructed into a five-dimensional complex-valued data cube $\mathbf{X} \in \mathbb{C}^{S \times N_{\mathrm{rx}} \times N_{\mathrm{tx}} \times L \times T}$, where $S{=}256$ denotes the number of fast-time ADC samples per chirp, $N_{\mathrm{rx}}{=}4$ and $N_{\mathrm{tx}}{=}3$ represent the number of receive and transmit antennas, respectively, $L{=}128$ corresponds to the number of chirps per frame, and $T$ indexes the consecutive temporal frames.

For mmWave-based classification and recognition tasks~\cite{shi2025human, santhalingam2020mmasl, sharma2023radar, alanazi2022towards}, the raw radar signals are typically transformed into Range--Doppler spectrograms for subsequent analysis.
Fig.~\ref{fig:radar_pipeline} illustrates the end-to-end radar signal processing pipeline implemented in our framework to extract the Range--Doppler spectrum.
To improve signal-to-noise ratio (SNR) and suppress background clutter and spectral leakage,
a multi-stage processing sequence combining windowing, filtering, and Fourier transformations is adopted, as detailed below.
\added{Standard mathematical definitions (e.g., range/Doppler FFT, MTI/mean removal, coherent integration, and the RDM dB conversion) are detailed in Appendix~\ref{app:rdm_formulas}.}

\begin{itemize}
\item \textbf{Range Processing:}
For each virtual aperture formed by the $t$-th transmit and $u$-th receive antenna pair,
a Hann tapering window is applied along the fast-time (ADC sample) dimension to reduce range sidelobes.
An $S$-point FFT converts the time-domain signal into the range-frequency domain, yielding $S$ range bins.
Under our radar configuration, the resulting range resolution is approximately $4.76$~cm.

\item \textbf{Clutter Suppression:}
Static reflections from walls, furniture, and other stationary objects are removed by subtracting
the temporal mean across the slow-time (chirp loop) dimension.
This mean-removal step functions as a first-order high-pass filter that suppresses near-zero Doppler components
while retaining dynamic targets with measurable radial velocity.

\item \textbf{Doppler Processing:}
After clutter removal, another Hann window is applied along the slow-time dimension to mitigate spectral leakage.
An $L$-point FFT is then performed to obtain Doppler frequency components.
The FFT-shift operation re-centers the DC component, producing a symmetric Doppler spectrum.
Under our configuration, the Doppler resolution is approximately $0.211$~m/s.

\item \textbf{MIMO Integration and RDM Formation:}
The Doppler-processed signals from all $N_{\mathrm{tx}} \times N_{\mathrm{rx}}{=}12$ virtual channels
are coherently combined (complex summation before magnitude computation)
to generate the final Range--Doppler Map (RDM). The resulting RDM is converted to the decibel (dB) scale with a small numerical stabilizer.
\end{itemize}

\subsubsection{Aligning RGB and Depth Frames from the Kinect Sensor.}

The RGB frames captured by the Kinect sensor have a native resolution of 1280×720 pixels. For spatial normalization, we first compute a global bounding box that encompasses all detected human keypoints across each activity sequence. Specifically, we determine the minimum and maximum 
$x$ and $y$ coordinates from all detected keypoints and expand the resulting bounding box by 10\% on each side to include contextual information while keeping the subject centered in the frame. All RGB frames are then center-cropped to this square region and resized to 224×224 pixels using bilinear interpolation, following standard ImageNet\cite{deng2009imagenet} preprocessing conventions to facilitate transfer learning from pretrained vision models.

The depth frames, originally captured at 640×576 pixels, are processed using a consistent spatial pipeline. We first center-crop the frames to 576×576 pixels (removing 32 pixels from each horizontal edge) to obtain square images, which are then resized to 224×224 pixels using nearest-neighbor interpolation to preserve the integrity of depth values. 

\subsubsection{Spectrum Calculation for IMUs Time-series.}
\label{sec:pre_imus}

Each wrist-worn IMU device continuously records raw tri-axial acceleration and angular velocity signals at a fixed sampling rate.
In our setup, two WT901 Wi-Fi IMUs are mounted on both wrists, yielding a total of 12 motion channels (6 per wrist), including 3-axis accelerometer and 3-axis gyroscope signals.
\myadded{All IMU signals are temporally aligned across devices and uniformly resampled to 50 Hz, ensuring synchronization between the two wrist sensors.
\added{Following prior work on wearable inertial sensing~\cite{chen2019taprint, chen2021vifin, wu2025vibrun}, we standardize the resampled sequences using per-channel z-score normalization to eliminate scale inconsistencies. The z-score normalization is applied individually to each clip, ensuring consistent scaling within each clip (Appendix~\ref{app:imu_formulas}).}}

To transform these time-domain signals into frequency-domain representations that align with other sensing modalities, we apply the Short-Time Fourier Transform (STFT) with a Hann window to each channel \added{(Appendix~\ref{app:imu_formulas})}.
The magnitude spectra are then log-compressed and normalized to enhance numerical stability and suppress dynamic range differences.
Finally, the spectrograms from all 12 channels are stacked along the channel dimension to form a three-dimensional tensor $\mathbf{X}_{imu}^{spec} \in \mathbb{R}^{T \times F \times C}$, where $T$ represents the number of time frames, $F$ is the frequency resolution, and $C=12$ corresponds to the number of IMU signal channels.

\subsubsection{Outliers Removal.}

At last, to ensure data quality, we remove outlier samples caused by synchronization failures during data collection. For each multimodal word clip, we compute the temporal deviation between modalities at both the start and end timestamps. A sample is classified as an outlier and excluded if the maximum temporal deviation exceeds 33~ms, corresponding to one frame period of the depth camera operating at 30~fps:
\begin{equation}
\max(\epsilon_{\text{start}}, \epsilon_{\text{end}}) > 33~\text{ms}
\end{equation}
where $\epsilon_{\text{start}}$ and $\epsilon_{\text{end}}$ denotes the temporal misalignment at segment boundaries. This threshold-based filtering strategy effectively eliminates poorly synchronized samples while preserving temporal coherence across modalities.

After applying the above preprocessing procedures, the \systemname{} dataset provides not only high-quality multimodal raw recordings but also processed word-level sign clips, including aligned RGB and depth frames, mmWave spectrograms, and IMU spectrograms, thereby supporting convenient and reproducible use of our large-scale dataset for future research.

\subsection{Dataset Statistics}

\begin{figure}[t]
    \centering
    \includegraphics[width=\linewidth]{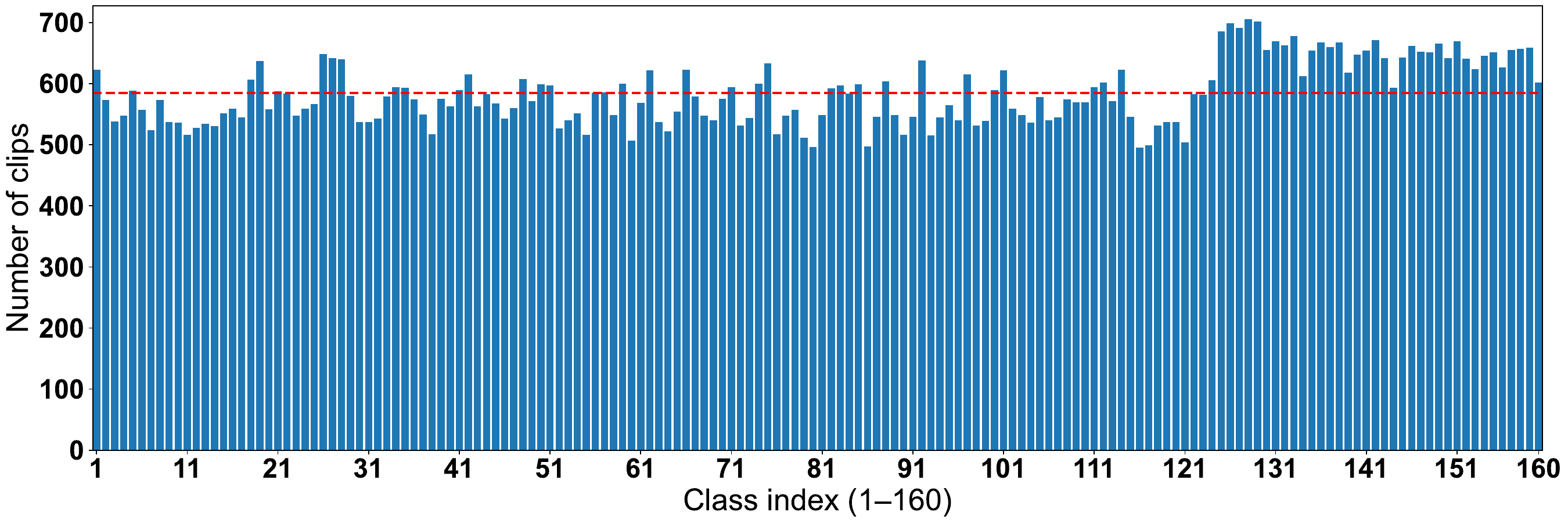}
    \caption{\myadded{Number of clips per class. This figure illustrates the distribution of the number of clips across different classes in the dataset. The red dashed line represents the average number of clips across all classes.}}
    \label{fig:app_clips_per_class}
\end{figure}

\begin{figure}[t]
    \centering
    \includegraphics[width=0.85\linewidth]{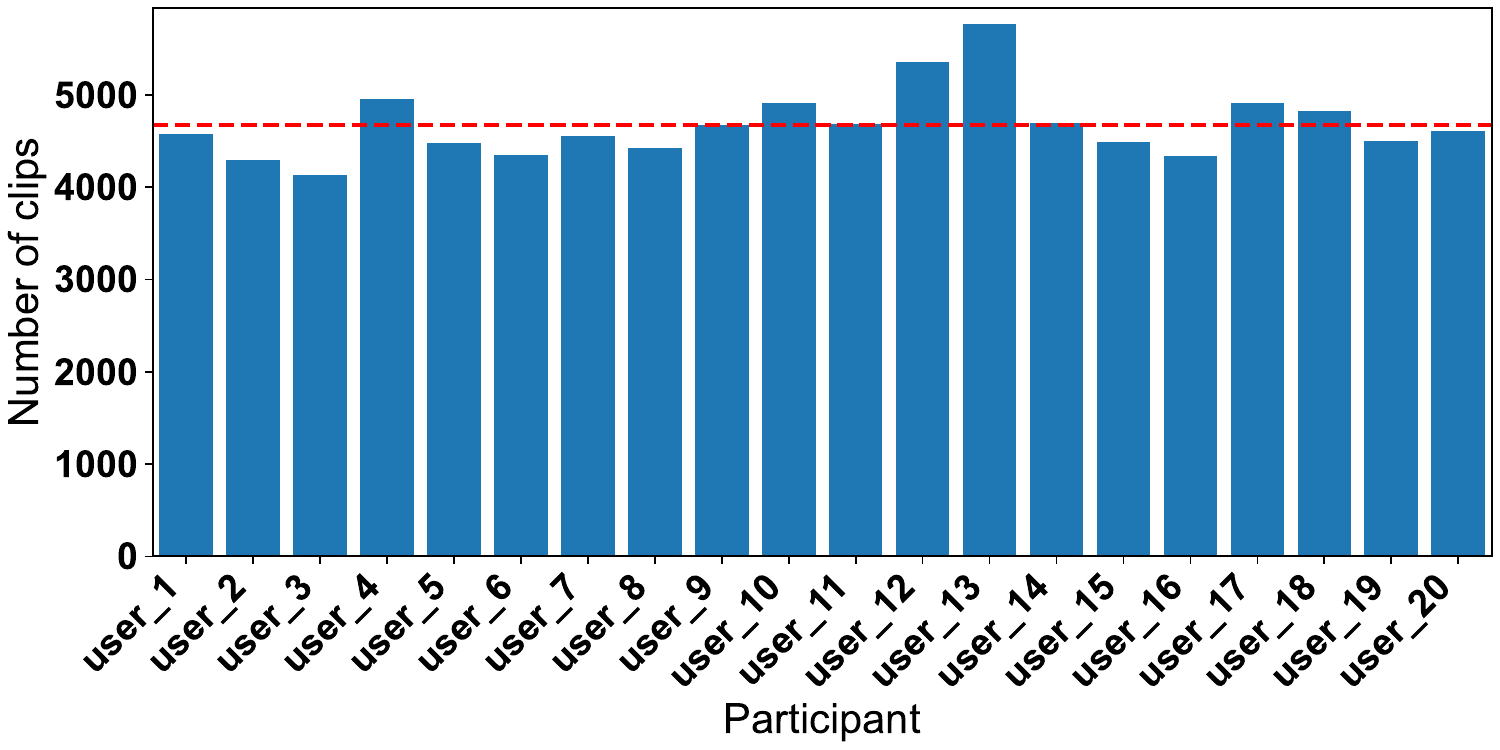}
    \caption{\myadded{Number of clips per participant. This figure shows the distribution of the number of clips across participants in the study. The red dashed line indicates the average number of clips across all participants.}}
    \label{fig:app_clips_per_user}
\end{figure}

Finally, we collected a large-scale multimodal dataset for ASL recognition. In total, the \systemname{} dataset contains approximately 6 million RGB frames, 6 million depth frames, 40 million IMU data points, and 6 million mmWave radar spatio-temporal spectrum frames. The raw data occupy more than 17 TB of storage, making \systemname{} one of the largest multimodal ISLR datasets to date.
To facilitate future research and reduce the effort required for dataset preparation, both the processed samples and the raw recordings will be publicly released.
After preprocessing, the dataset consists of \added{93,545} multimodal word clips contributed by 20 participants, each of whom devoted over 16 effective hours to data collection. On average, each participant contributed approximately \added{4,677} word clips recorded across two independent sessions. \added{Specifically, the per-class clip counts range from \added{495} to \added{706} across \added{160} classes, with a mean of \added{584.66} and a Coefficient of Variation (CV, Appendix \ref{app:cv_definition}) of \added{0.086}. The per-participant clip counts range from \added{4,132} to \added{5,766} across \added{20} participants, with a mean of \added{4,677.25} and a CV of \added{0.079}. As shown in Figures \ref{fig:app_clips_per_class} and \ref{fig:app_clips_per_user}, these distributions indicate that the dataset is reasonably balanced without severe imbalances or long-tail sparsity.} Each vocabulary item has at least 400 valid and synchronized clips collected across all participants, ensuring sufficient diversity and statistical robustness for model training and evaluation.

\begin{figure}[h]
  \centering
  \begin{subfigure}[t]{0.65\linewidth}
    \centering
    \includegraphics[width=\linewidth]{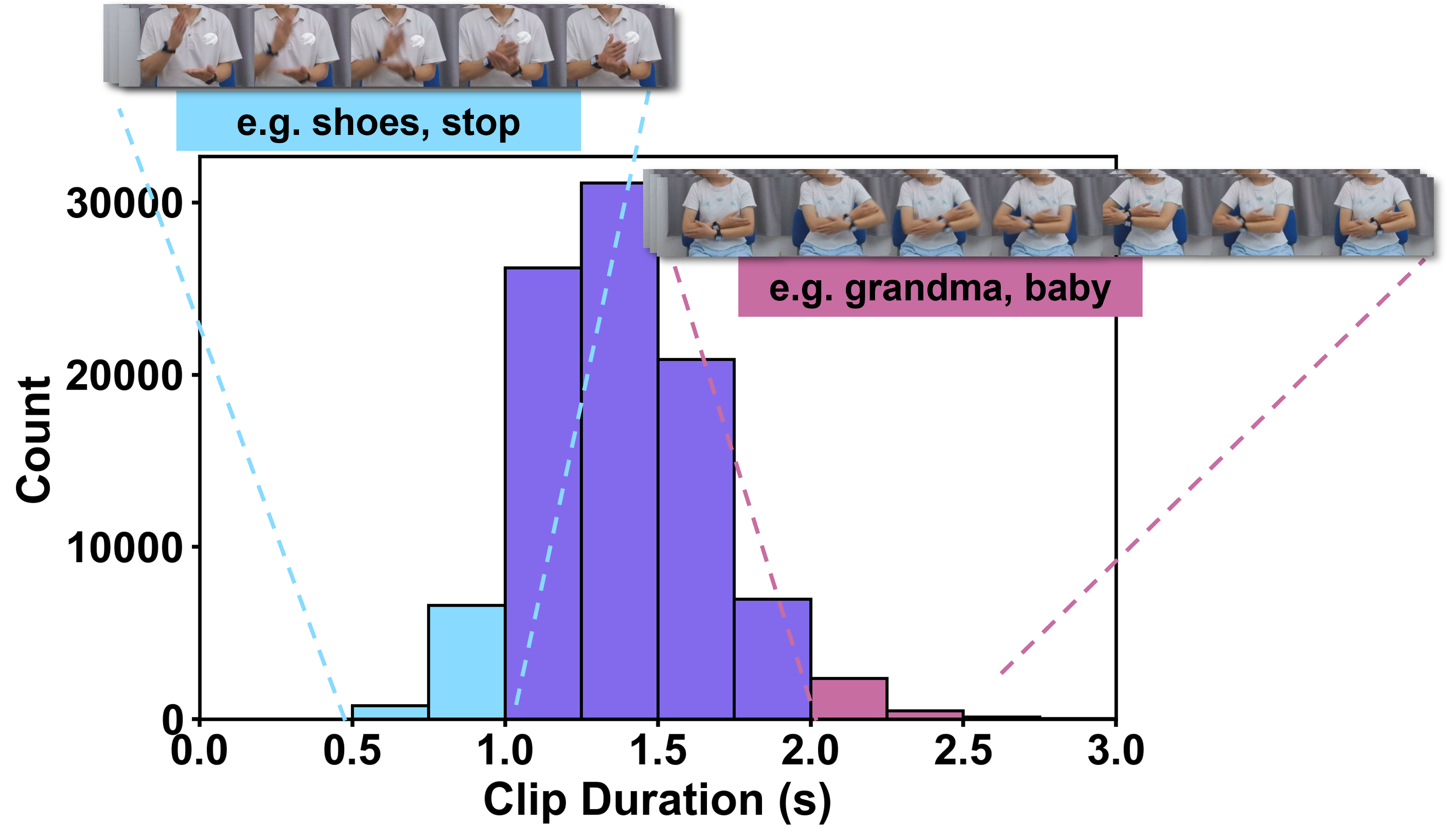}
    \subcaption{Duration distribution of word clips.}
    \label{fig:clip_duration}
  \end{subfigure}
  \hfill
  \begin{subfigure}[t]{0.65\linewidth}
    \centering
    \includegraphics[width=\linewidth]{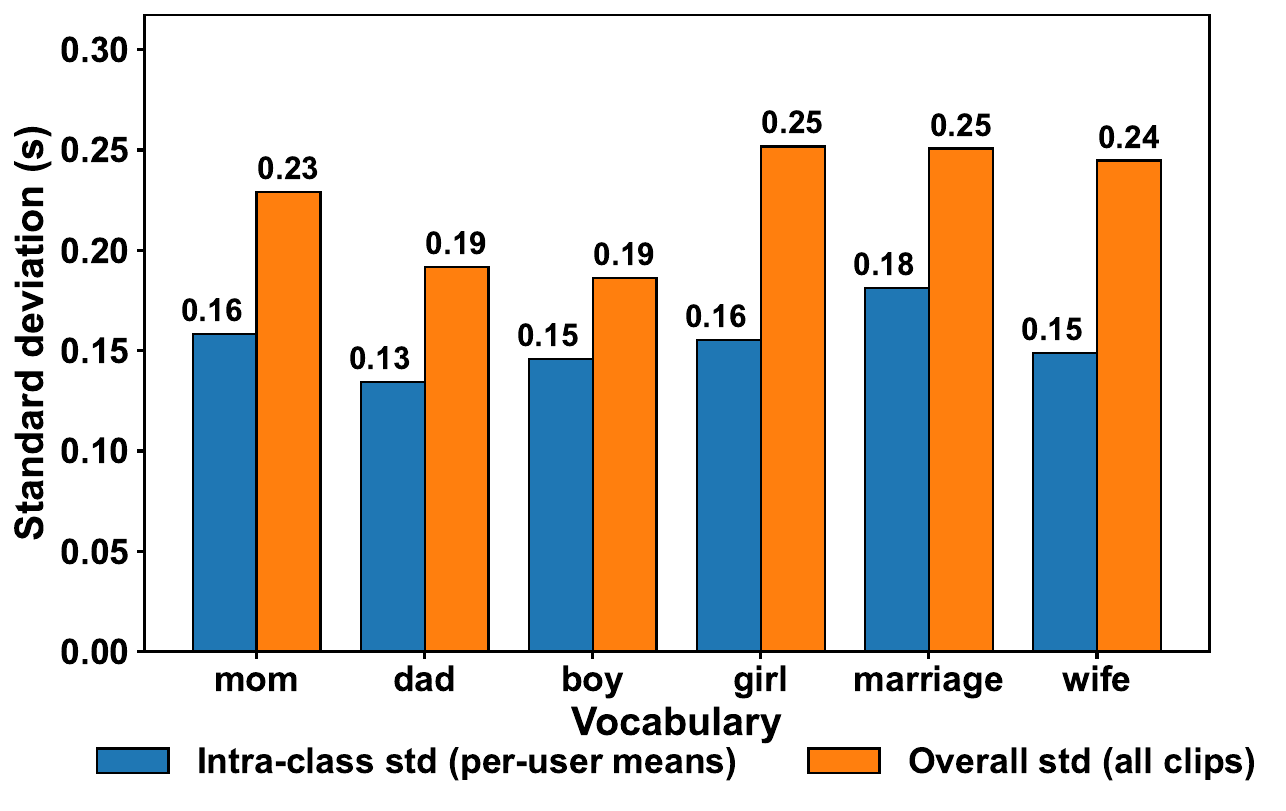}
    \subcaption{Comparison of intra- vs.\ inter-signer temporal variability.}
    \label{fig:intra_inter_std}
  \end{subfigure}

  \caption{
  (a) Distribution of word-clip durations in the \systemname{} dataset.
  Nearly all signs fall within the 0--3\,s range, with most concentrated between 1.0--2.0\,s, accounting for approximately 90\% of all samples.
  (b) Comparison of intra-signer (blue) and inter-signer (orange) standard deviations of clip durations for ten representative vocabulary items. 
  The consistently higher orange bars indicate that inter-signer temporal variability (differences across participants) is substantially greater than intra-signer variability (fluctuations within the same participant).}
  \label{fig:clip_duration_combined}
\end{figure}

Due to the varying complexity of ASL vocabulary and the diverse signing speeds among participants, the durations of recorded word clips exhibit substantial variation. Fig.~\ref{fig:clip_duration} presents the overall distribution of clip durations. Nearly all samples fall within the 0–3 seconds range, with most concentrated between 1.0 and 2.0 seconds, accounting for approximately 90\% of all clips.
Simple, single-movement signs (e.g., shoes, stop) typically last less than one second, whereas more complex or multi-gesture signs (e.g., grandma, baby) often exceed two seconds. This variability reflects differences in motion complexity and individual signing rhythm across participants.

To further quantify temporal consistency across participants, we computed the standard deviation of clip durations for each vocabulary item both within individual signers and across all signers, as illustrated in Fig.~\ref{fig:intra_inter_std}. Here, ten representative vocabulary items were randomly selected for visualization. The blue bars represent the intra-signer standard deviation, indicating the average temporal fluctuation when the same signer repeatedly performs the same word, while the orange bars denote the overall standard deviation, reflecting the global variation across all signers. It can be observed that, for all vocabulary items, the inter-signer variability (differences across participants) is substantially greater than intra-signer variability (temporal fluctuation within individual signers). This indicates that each signer tends to perform a given sign with stable timing patterns, whereas variations across participants lead to greater overall diversity. These findings underscore the importance of adopting \textbf{user-independent }evaluation protocols in sign language recognition and highlight the need for models with strong cross-signer robustness to handle temporal variations arising from individual differences.

\section{Benchmarking Methodologies}

In this section, we briefly describe the benchmarking methods adopted to assess the SLR accuracy on the \systemname{} dataset. Since the dataset contains multiple sensing modalities, the benchmarking protocols include both single-modality and multimodality approaches.

\subsection{Single-modality Benchmarking Methods}
The ISLR task can be performed using any single modality collected in \systemname{}. Based on the sensing modality utilized, the benchmarking methods can be broadly classified into {\bf vision-based}, {\bf mmWave-based}, and {\bf IMU-based} ISLR approaches.
\subsubsection{Vision-based ISLR}
The visual modality, comprising both RGB and depth frames, plays a central role in contemporary action and sign language recognition research. To establish robust visual baselines on the \systemname{} dataset, we adopt several state-of-the-art architectures capable of processing RGB, depth, or combined RGB-D inputs for isolated sign language recognition. Specifically, we evaluate five representative models: I3D~\cite{carreira2017quo}, TimeSformer~\cite{bertasius2021space}, SlowFast~\cite{feichtenhofer2019slowfast}, S3D~\cite{xie2018rethinking}, and UMDR~\cite{zhou2023unified}:

\begin{itemize}
\item {\bf I3D}~\cite{carreira2017quo} models joint spatiotemporal representations using inflated 3D convolutions, enabling simultaneous learning of motion and appearance features.
\item {\bf TimeSformer}~\cite{bertasius2021space} introduces a transformer-based architecture with divided space–time attention, effectively capturing long-range temporal dependencies.
\item {\bf SlowFast}~\cite{feichtenhofer2019slowfast} adopts a dual-pathway design to separately model semantic context at low frame rates and motion dynamics at high frame rates.
\item {\bf S3D}~\cite{xie2018rethinking} utilizes separable 3D convolutions to reduce computational complexity while preserving temporal fidelity in motion representation.
\item {\bf UMDR}~\cite{zhou2023unified} presents a unified multimodal de- and re-coupling framework that enhances spatiotemporal feature learning for RGB-D motion understanding.
\end{itemize}

All models described above are trained under a consistent experimental protocol using either 32-frame or 64-frame clips with a spatial resolution of $224 \times 224$ pixels. The RGB and Depth modalities are processed independently but share identical preprocessing and temporal sampling strategies to ensure fair comparability. 
Notably, {\bf I3D} and {\bf TimeSformer} are initialized with pretrained weights from the Kinetics-400 dataset~\cite{kay2017kinetics} to leverage large-scale spatiotemporal priors and accelerate training convergence.

\subsubsection{mmWave-based ISLR}

mmWave radar provides a complementary sensing modality for sign language recognition by capturing fine-grained motion and velocity information through radio-frequency reflections. 
In \systemname{}, radar data are processed into time-sequential Range-Doppler Maps (RDMs), representing reflected signal energy across distance and velocity domains. 
To the best of our knowledge, no established benchmarks exist specifically for mmWave-based ISLR. However, prior work has demonstrated the feasibility of radar for hand gesture and sign identification~\cite{santhalingam2020expressive,lan2023applying,gurbuz2020american}. Following these insights, we implement several representative methods as radar baselines:
\begin{itemize}
\item \textbf{CNN+BiGRU} combines convolutional spatial feature extraction with bidirectional temporal modeling.
\item \textbf{CNN+LSTM} employs CNNs for spatial encoding followed by LSTM for sequential temporal dependencies.
\item \textbf{Transformer} captures long-range dependencies through multi-head self-attention mechanisms.
\item \textbf{I3D} directly extracts spatiotemporal representations from stacked RDM sequences via 3D convolutions.
\end{itemize}

\subsubsection{IMU-based ISLR}

IMUs offer a compact, motion-centric, and privacy-preserving modality for sign language recognition. They can accurately capture subtle wrist and hand dynamics under challenging conditions such as poor lighting, occlusion, or privacy-sensitive environments. In \systemname{}, we provide both raw multi-axis time-series data and their corresponding time–frequency representations derived via short-time Fourier transform (STFT). Based on these two forms of input, we implement a representative set of IMU-based ISLR architectures that reflect both temporal sequence modeling and time–frequency analysis paradigms.

\begin{itemize}
\item \textbf{TCN} leverages dilated convolutions to model long-range temporal dependencies in sequential motion data.
\item \textbf{Lightweight Transformer} applies efficient self-attention to capture global temporal correlations across IMU channels.
\item \textbf{BiLSTM-Attn} integrates bidirectional recurrence with attention-based feature aggregation for adaptive temporal weighting.
\item \textbf{CNN-GRU} combines local convolutional feature extraction with gated recurrent units for temporal dynamics modeling.
\item \textbf{I3D (Spectrogram)} extends spatiotemporal convolutional learning to time–frequency representations of IMU signals (obtained from preprocessing on IMU time-series described in Sec. \ref{sec:pre_imus}), effectively capturing rhythmic and oscillatory patterns.
\end{itemize}

\subsection{Multi-modality Benchmarking Methods}

Multimodal fusion aims to integrate the complementary information provided by heterogeneous sensors, thereby constructing a more comprehensive and robust representation of sign language actions. In the multimodal fusion benchmark of \systemname{}, RGB, Depth, mmWave Radar, and IMU modalities all employ \textbf{I3D} as the backbone encoder and undergo independent single-modality pre-training. During fusion, a partial-freeze strategy is adopted: the first four feature extraction stages are frozen, while only the top-level feature stage and the final classification layer are fine-tuned. This design preserves generic spatiotemporal representations while substantially reducing the risks of overfitting and the overall training cost.
To address the inconsistency in confidence and numerical scale across modalities, a learnable calibration layer is applied to the temporal logits of each modality, performing temperature scaling and class-wise affine normalization~\cite{ding2021local}. Before calibration and fusion, the temporal lengths of modalities are aligned through linear interpolation to a unified number of frames. Fusion is conducted at the logits level using learnable modality weights that are normalized by Softmax, enabling adaptive contribution from each modality. To further stabilize the optimization of the multi-branch architecture, auxiliary cross-entropy losses are imposed on the calibrated segment-level logits and jointly optimized with the main loss. The final prediction is obtained by temporal averaging of the fused logits. Throughout the entire training process, only the fusion head, the modality-wise calibration layers, and the unfrozen top layers of the \textbf{I3D} backbones are updated, ensuring stable optimization and accelerated convergence.

\section{Evaluation on \systemname{} Dataset}
\label{sec:validation}

In this section, we present comprehensive benchmarking results to demonstrate the capabilities and challenges of the dataset across multiple sensing modalities. We establish systematic baselines by evaluating the four distinct modalities and their combinations under two evaluation protocols: user-independent (UI) and user-dependent (UD), which evaluates cross-signer generalization and cross-session robustness.

\subsection{Evaluation Metric and Protocols}
\subsubsection{Evaluation Metric.}

We adopt Top-$k$ Accuracy as the core evaluation metric to assess benchmarking methods on the \systemname{} dataset. This metric measures how often the correct label appears within the model's top-$k$ ranked predictions, making it well-suited for ISLR tasks that involve a large number of candidate classes \cite{li2020word,shen2024mm,joze2018ms,shen2023auslan}. \added{The standard mathematical definition of Top-$k$ Accuracy is provided in Appendix~\ref{app:eval_formulas} (Eq.~\ref{eq:topk}).}

In our evaluation, we report Top-1, Top-3, and Top-5 accuracies to assess model performance under different configurations. The Top-1 accuracy measures the model’s confidence in its primary prediction and serves as the main metric for comparison. Top-3 and Top-5 accuracies provide additional insights into model robustness, as the correct label often appears among the top-ranked predictions even when it is not the highest ranked. These metrics are particularly valuable in practical applications where downstream systems can leverage contextual or semantic information, such as that provided by LLMs~\cite{pradeep2023rankvicuna,jin2024llm,nam2024using}, to refine or re-rank recognition results.

\subsubsection{Evaluation Protocols.} 
We evaluate the \systemname{} dataset under two protocols to assess different aspects of model generalization:
\begin{itemize}
\item{\bf User-dependent (UD) Evaluation:} Training data are collected from the first session of participants 1–12, while test data are collected from the second session of the same participants. This protocol evaluates \textit{cross-session robustness} by assessing whether models can generalize across different recording sessions for known users.

\item{\bf User-independent (UI) Evaluation:} Training and validation data are collected from the first session of participants 1–12, while test data are collected from the remaining participants. The test set contains entirely unseen signers, creating an unseen scenario that evaluates \textit{cross-signer generalization} without user-specific fine-tuning. This protocol is more challenging and practical, as it reflects real-world deployment where the system must recognize signs from new users.
\end{itemize}

\subsection{Single-Modality Evaluation Results}
\label{sec:single_validaition}

\begin{table*}[htb]
  \centering
  \caption{Single-modality benchmarks on \systemname{} under \textbf{UI} and \textbf{UD} protocols. Each entry reports Top-1 / Top-3 / Top-5 accuracies (\%). The highest accuracy is marked in {\bf bold}.}
  \label{tab:single_modality_ui_ud}
  \begin{tabular}{ll|ccc|ccc}
    \toprule
    \multirow{2}{*}{\textbf{Modality}} & 
    \multirow{2}{*}{\textbf{Methods}} &
    \multicolumn{3}{c|}{\textbf{User-Independent (UI)}} &
    \multicolumn{3}{c}{\textbf{User-Dependent (UD)}} \\
    \cmidrule(lr){3-5}\cmidrule(lr){6-8}
     & & \textbf{Top-1} & \textbf{Top-3} & \textbf{Top-5} & 
       \textbf{Top-1} & \textbf{Top-3} & \textbf{Top-5} \\
    \midrule
    \multirow{5}{*}{IMU}
      & TCN                       & 55.08 & 71.27 & 76.20 & 64.94 & 79.44 & 83.88 \\
      & Lightweight Transformer   & 49.65 & 65.10 & 71.38 & 59.02 & 75.28 & 81.53 \\
      & BiLSTM-Attn               & 57.53 & 70.78 & 75.70 & 67.22 & 79.33 & 83.39 \\
      & CNN-GRU                   & 62.31 & 75.16 & 79.32 & \textbf{71.29} & \textbf{83.12} & \textbf{86.81} \\
      & I3D (STFT)                & \textbf{63.10} & \textbf{77.58} & \textbf{82.08} & 70.37 & 82.43 & 86.44 \\
    \midrule
    \multirow{4}{*}{mmWave}
      & CNN+BiGRU                 & 50.30 & 68.46 & 74.92 & 57.57 & 74.78 & 80.36 \\
      & CNN+LSTM                  & 46.35 & 65.33 & 72.52 & 50.90 & 69.98 & 76.38 \\
      & Lightweight Transformer   & 46.37 & 65.01 & 71.74 & 51.73 & 69.69 & 75.73 \\
      & I3D                       & \textbf{70.92} & \textbf{84.82} & \textbf{89.47} & \textbf{79.58} & \textbf{90.76} & \textbf{94.17} \\
    \midrule
    \multirow{5}{*}{RGB}
      & I3D                       & 88.12 & 96.76 & 98.29 & 93.47 & 99.01 & 99.58 \\
      & TimeSformer               & 80.31 & 92.26 & 95.03 & 90.80 & 98.24 & 99.16 \\
      & SlowFast                  & 88.59 & 96.52 & 97.84 & 94.92 & 99.03 & 99.39 \\
      & S3D                       & 49.88 & 73.62 & 82.62 & 57.54 & 81.72 & 89.38 \\
      & UMDR-M                    & \textbf{92.68} & \textbf{98.08} & \textbf{98.81} & \textbf{96.21} & \textbf{99.44} & \textbf{99.69} \\
    \midrule
    \multirow{4}{*}{Depth}
      & I3D                       & 76.45 & 91.04 & 94.57 & 85.03 & 95.47 & 97.71 \\
      & TimeSformer               & 71.30 & 88.96 & 93.43 & 81.99 & 95.16 & 97.81 \\
      & SlowFast                  & 87.36 & 95.50 & 97.29 & 93.49 & 98.63 & 99.30 \\
      & UMDR-K                    & \textbf{91.17} & \textbf{98.01} & \textbf{98.91} & \textbf{94.86} & \textbf{99.30} & \textbf{99.67} \\
    \bottomrule
  \end{tabular}

  \vspace{3pt}
\end{table*}

We establish baselines by training separate models for each modality to assess their individual capabilities: RGB, Depth, IMU, and mmWave Radar. All models use identical train-validation splits (12:8 participants) to ensure fair comparison across modalities. Table~\ref{tab:single_modality_ui_ud} reports Top-1/Top-3/Top-5 accuracies under both UD and UI protocols, revealing several key findings:

\noindent\textbf{Visual vs. Non-Visual Modalities.} RGB and Depth significantly outperform IMU and mmWave across both evaluation settings. RGB achieves the highest accuracy (UI: 92.71\% with UMDR-M; UD: 94.92\% with SlowFast), followed closely by Depth (UI: 91.21\% with UMDR-K; UD: 93.49\% with SlowFast). In contrast, non-visual modalities show substantially lower accuracy: IMU peaks at 63.10\% (UI) and mmWave at 70.92\% (UI). Despite lower accuracy, IMU and mmWave offer potential practical advantages for privacy-sensitive deployments and challenging environmental conditions, such as the ability to operate in darkness, through occlusions, and without capturing personally identifiable information.

\noindent\textbf{UI vs. UD Performance Gap.} All modalities show consistent performance improvements from UI to UD settings, with an average gain of 8-10 percentage points in Top-1 accuracy. For example, RGB TimeSformer improves from 80.31\% (UI) to 90.80\% (UD). 
This gap underscores a critical limitation in current ISLR approaches: the ability to adapt to known users' temporal variations does not translate to robustness against inter-user variability, revealing the need for more user-invariant representations.

\noindent\textbf{Methodology Comparisons.} Within each modality, I3D and SlowFast consistently outperform RNN-based and Transformer-based methods. For IMU, I3D with STFT preprocessing achieves 63.10\% (UI), outperforming CNN-GRU by 0.79\%. For mmWave, I3D reaches 70.92\% (UI), substantially exceeding CNN+BiGRU (50.30\%) by over 20\%. This suggests that 3D convolutional architectures of I3D better capture spatiotemporal patterns for sign recognition across diverse sensing modalities.

\noindent\textbf{Top-k Metric Insights.} Top-3 and Top-5 accuracies reveal significant potential for improvement through contextual post-processing. For example, the RGB-based I3D model achieves 96.76\% Top-3 accuracy under user-independent evaluation, despite an 88.12\% Top-1 score, indicating that the correct sign frequently appears among the top-ranked predictions. This trend is consistent across all sensing modalities, suggesting that practical systems could further enhance recognition performance by incorporating contextual reasoning or language models to disambiguate among the most probable candidates.

\noindent\textbf{Modality Complementarity.}

\begin{table}[htb]
\centering
\caption{\myadded{Top-20 hardest words to be recognized across RGB, IMU, and mmWave modalities. Reported values are Top-1 accuracy, with boldface indicating that other modalities outperform the current modality on the corresponding word.}}
\label{tab:hardest_combined}
\begin{adjustbox}{width=\textwidth} %
\small
\setlength{\tabcolsep}{3pt}
\renewcommand{\arraystretch}{1.2}
\begin{tabular}{llcccccccccccccccccccc}
\toprule
\textbf{} & \textbf{Modality} & \multicolumn{20}{c}{\textbf{Top-20 Hardest Words}} \\
\midrule
\multirow{4}{*}{\rotatebox[origin=c]{90}{\textbf{RGB}}} & word & good & t & w & nine & six & n & i & eight & f & s & milk & apple & dog & bug & o & g & excuse & what\textsuperscript{1} & a & r \\
\cmidrule{2-22}
& \cellcolor{gray!20}RGB(\%) & \cellcolor{gray!20}14.9 & \cellcolor{gray!20}36.0 & \cellcolor{gray!20}37.0 & \cellcolor{gray!20}40.0 & \cellcolor{gray!20}44.5 & \cellcolor{gray!20}46.2 & \cellcolor{gray!20}55.9 & \cellcolor{gray!20}58.1 & \cellcolor{gray!20}59.0 & \cellcolor{gray!20}59.8 & \cellcolor{gray!20}59.8 & \cellcolor{gray!20}61.1 & \cellcolor{gray!20}62.7 & \cellcolor{gray!20}62.9 & \cellcolor{gray!20}66.2 & \cellcolor{gray!20}66.9 & \cellcolor{gray!20}66.9 & \cellcolor{gray!20}67.5 & \cellcolor{gray!20}71.5 & \cellcolor{gray!20}72.1 \\
& IMU(\%) & \textbf{18.1} & 10.4 & 9.4 & 6.4 & 2.1 & 0.8 & 0.0 & 3.5 & 3.1 & 10.2 & \textbf{80.2} & 39.8 & 54.2 & 29.3 & 4.4 & 24.8 & \textbf{74.3} & \textbf{81.7} & 9.2 & 29.5 \\
& mmWave(\%) & \textbf{45.8} & 12.0 & \textbf{40.2} & 12.9 & 11.0 & 19.7 & 5.5 & 10.6 & 9.4 & 19.7 & \textbf{62.7} & \textbf{67.0} & 57.6 & 50.9 & 12.5 & 18.8 & \textbf{97.5} & \textbf{78.3} & 35.4 & 10.9 \\
\midrule

\multirow{4}{*}{\rotatebox[origin=c]{90}{\textbf{IMU}}} & word & i & n & c & u & y & six & f & e & eight & o & l & nine & seven & a & w & s & t & four & green & m \\
\cmidrule{2-22}
& RGB(\%) & \textbf{55.9} & \textbf{46.2} & \textbf{100.0} & \textbf{78.8} & \textbf{77.1} & \textbf{44.5} & \textbf{59.1} & \textbf{81.2} & \textbf{58.2} & \textbf{66.2} & \textbf{83.9} & \textbf{40.0} & \textbf{81.0} & \textbf{71.5} & \textbf{37.0} & \textbf{59.8} & \textbf{36.0} & \textbf{90.1} & \textbf{81.9} & \textbf{75.0} \\
& \cellcolor{gray!20}IMU(\%) & \cellcolor{gray!20}0.0 & \cellcolor{gray!20}0.8 & \cellcolor{gray!20}1.5 & \cellcolor{gray!20}1.5 & \cellcolor{gray!20}1.5 & \cellcolor{gray!20}2.1 & \cellcolor{gray!20}3.1 & \cellcolor{gray!20}3.4 & \cellcolor{gray!20}3.5 & \cellcolor{gray!20}4.4 & \cellcolor{gray!20}5.8 & \cellcolor{gray!20}6.4 & \cellcolor{gray!20}7.7 & \cellcolor{gray!20}9.2 & \cellcolor{gray!20}9.4 & \cellcolor{gray!20}10.2 & \cellcolor{gray!20}10.4 & \cellcolor{gray!20}10.5 & \cellcolor{gray!20}12.4 & \cellcolor{gray!20}12.5 \\
& mmWave(\%) & \textbf{5.5} & \textbf{19.7} & \textbf{39.8} & \textbf{15.9} & \textbf{8.4} & \textbf{11.0} & \textbf{9.4} & \textbf{52.1} & \textbf{10.6} & \textbf{12.5} & \textbf{26.3} & \textbf{12.9} & \textbf{16.2} & \textbf{35.4} & \textbf{40.2} & \textbf{19.7} & \textbf{12.0} & \textbf{21.7} & \textbf{16.2} & \textbf{26.6} \\
\midrule

\multirow{4}{*}{\rotatebox[origin=c]{90}{\textbf{mmWave}}} & word & i & y & f & eight & r & six & t & v & o & nine & three & u & green & seven & one & g & s & n & four & x \\
\cmidrule{2-22}
& RGB(\%) & \textbf{55.9} & \textbf{77.1} & \textbf{59.1} & \textbf{58.2} & \textbf{72.1} & \textbf{44.5} & \textbf{36.0} & \textbf{86.9} & \textbf{66.2} & \textbf{40.0} & \textbf{78.9} & \textbf{78.8} & \textbf{81.9} & \textbf{81.0} & \textbf{87.5} & \textbf{66.9} & \textbf{59.8} & \textbf{46.2} & \textbf{90.1} & \textbf{85.0} \\
& IMU(\%) & 0.0 & 1.5 & 3.1 & 3.5 & \textbf{29.5} & 2.1 & 10.4 & \textbf{13.1} & 4.4 & 6.4 & \textbf{15.6} & 1.5 & 12.4 & 7.7 & 13.3 & \textbf{24.8} & 10.2 & 0.8 & 10.5 & \textbf{52.8} \\
& \cellcolor{gray!20}mmWave(\%) & \cellcolor{gray!20}5.5 & \cellcolor{gray!20}8.4 & \cellcolor{gray!20}9.4 & \cellcolor{gray!20}10.6 & \cellcolor{gray!20}10.9 & \cellcolor{gray!20}11.0 & \cellcolor{gray!20}12.0 & \cellcolor{gray!20}12.3 & \cellcolor{gray!20}12.5 & \cellcolor{gray!20}12.9 & \cellcolor{gray!20}14.3 & \cellcolor{gray!20}15.9 & \cellcolor{gray!20}16.2 & \cellcolor{gray!20}16.2 & \cellcolor{gray!20}18.0 & \cellcolor{gray!20}18.8 & \cellcolor{gray!20}19.7 & \cellcolor{gray!20}19.7 & \cellcolor{gray!20}21.7 & \cellcolor{gray!20}22.8 \\
\bottomrule
\end{tabular}
\end{adjustbox}
\end{table}

\added{As shown in Tables~\ref{tab:hardest_combined}, we report the top-20 hardest-to-distinguish classes for RGB, IMU, and mmWave modalities. IMU and mmWave exhibit significant difficulty in recognizing finger-spelling gestures such as "i," "n," and "c", which involve minimal finger articulation and highly similar wrist orientations. As shown in Table~\ref{tab:hardest_combined} and visualized in Figure~\ref{tab:rgb-imu}, "i" and "n" are single-hand gestures with nearly identical IMU signals from both hands, resulting in frequent confusion for IMU-based methods.}

\added{In contrast, RGB struggles with gestures such as "excuse" and "what\textsuperscript{1}", where visual appearances are highly similar. However, the corresponding left- and right-hand IMU signals exhibit distinct motion patterns (Figure~\ref{tab:rgb-imu}), enabling reliable differentiation. These results highlight the complementary nature of RGB, IMU, and mmWave modalities: while IMU and mmWave are limited in capturing fine finger movements, they provide discriminative motion cues for gestures that are visually ambiguous. Combining modalities therefore improves robustness and recognition accuracy beyond single modality can achieve.}

\newcolumntype{C}[1]{>{\centering\arraybackslash}m{#1}}
\setlength{\tabcolsep}{2pt} %
\begin{figure*}[t]

\centering
\begin{tabular}{C{0.08\textwidth} C{0.38\textwidth} C{0.26\textwidth} C{0.26\textwidth}}
\toprule
Word & RGB sequence & Left-hand IMU & Right-hand IMU \\
\midrule

\begin{minipage}[t]{\linewidth}
    \centering
        \includegraphics[width=\linewidth, height=2.2cm, keepaspectratio]{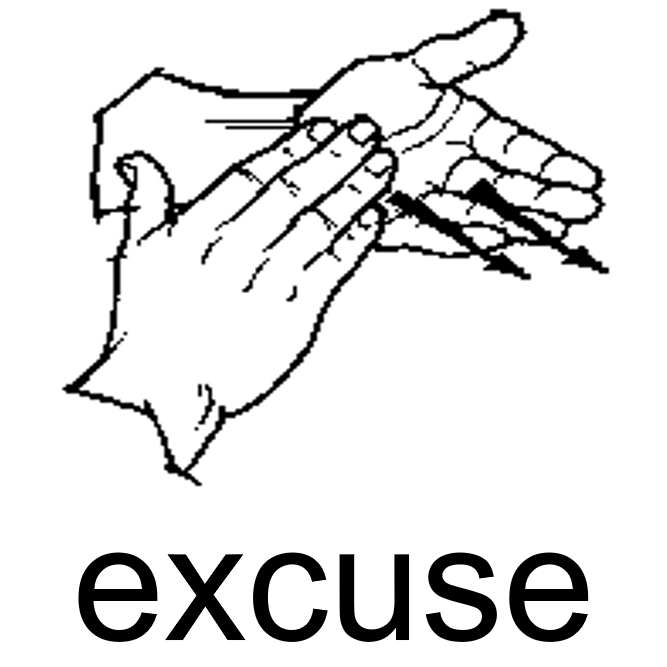}
\end{minipage}
&
\begin{minipage}[t]{\linewidth}
    \centering
    \includegraphics[width=\linewidth, height=2.2cm, keepaspectratio]{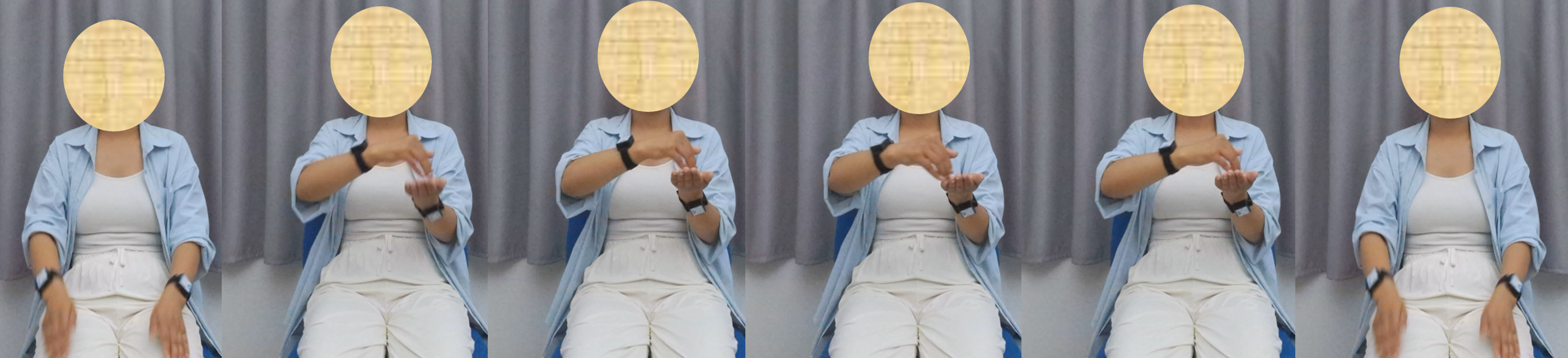}
\end{minipage}
&
\begin{minipage}[t]{\linewidth}
    \centering
    \includegraphics[width=\linewidth, height=2.2cm, keepaspectratio]{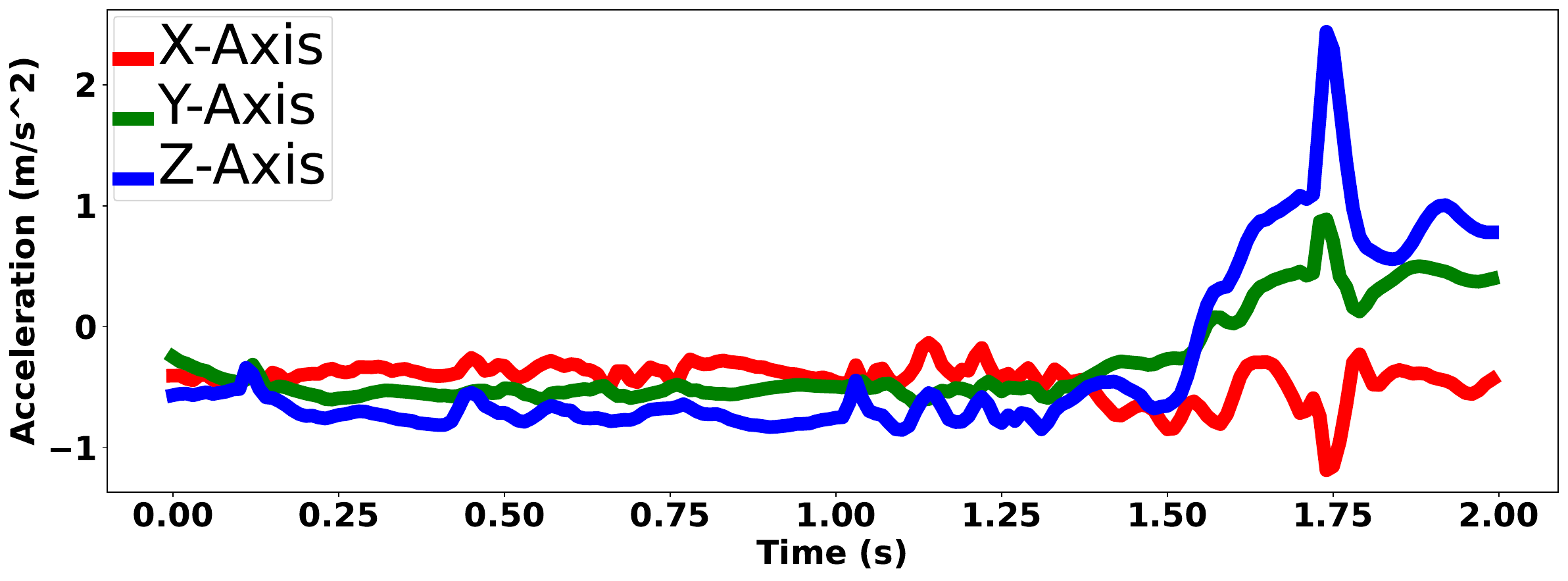}
\end{minipage}
&
\begin{minipage}[t]{\linewidth}
    \centering
    \includegraphics[width=\linewidth, height=2.2cm, keepaspectratio]{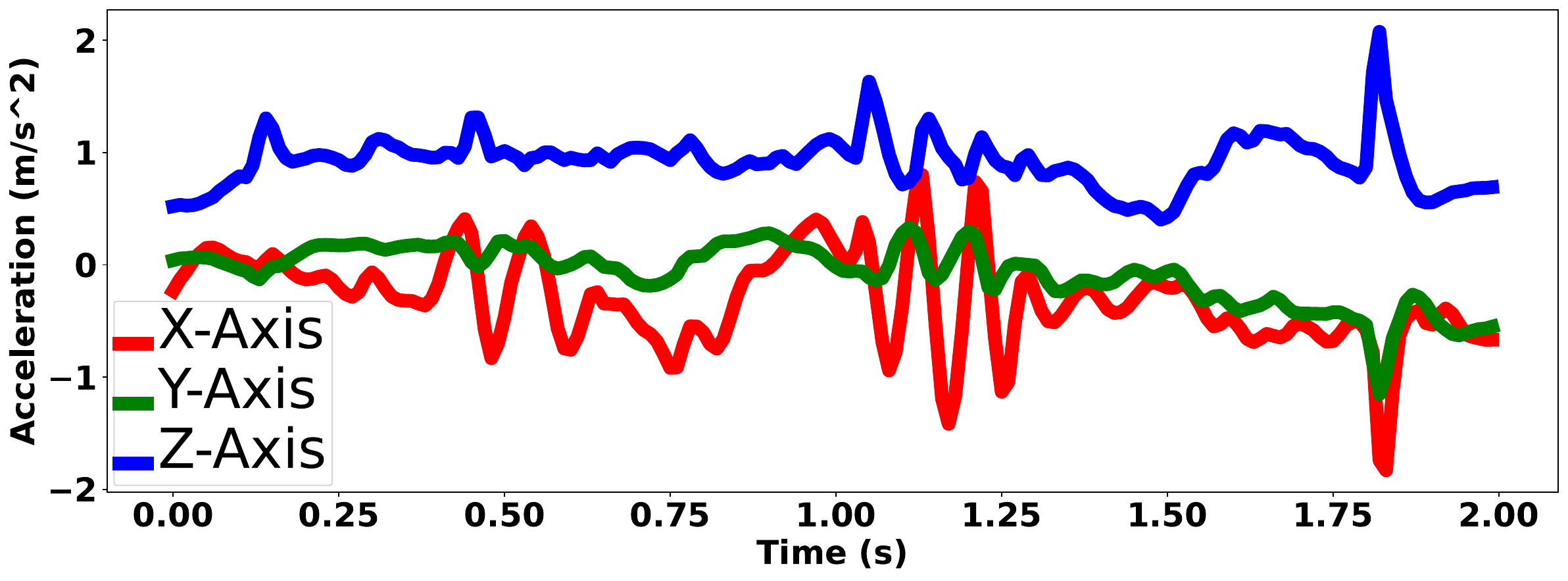}
\end{minipage}
\\

\begin{minipage}[t]{\linewidth}
    \centering
    \includegraphics[width=\linewidth, height=2.2cm, keepaspectratio]{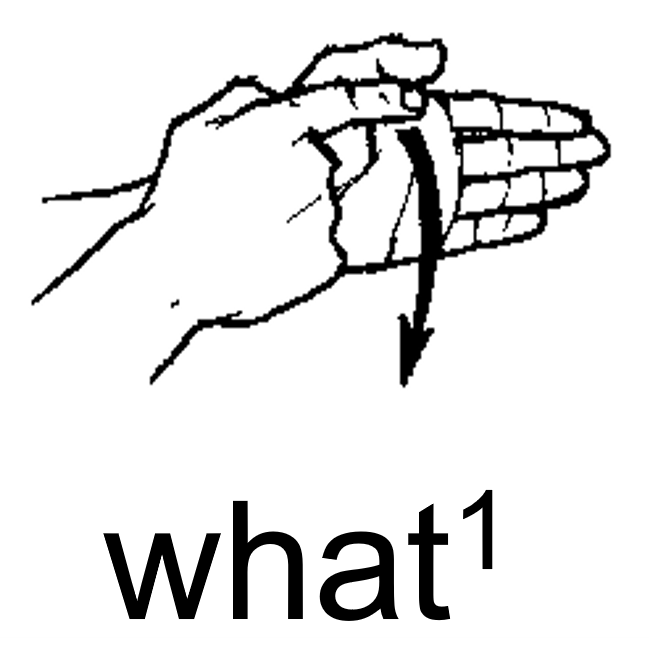}
\end{minipage}
&
\begin{minipage}[t]{\linewidth}
    \centering
    \includegraphics[width=\linewidth, height=2.2cm, keepaspectratio]{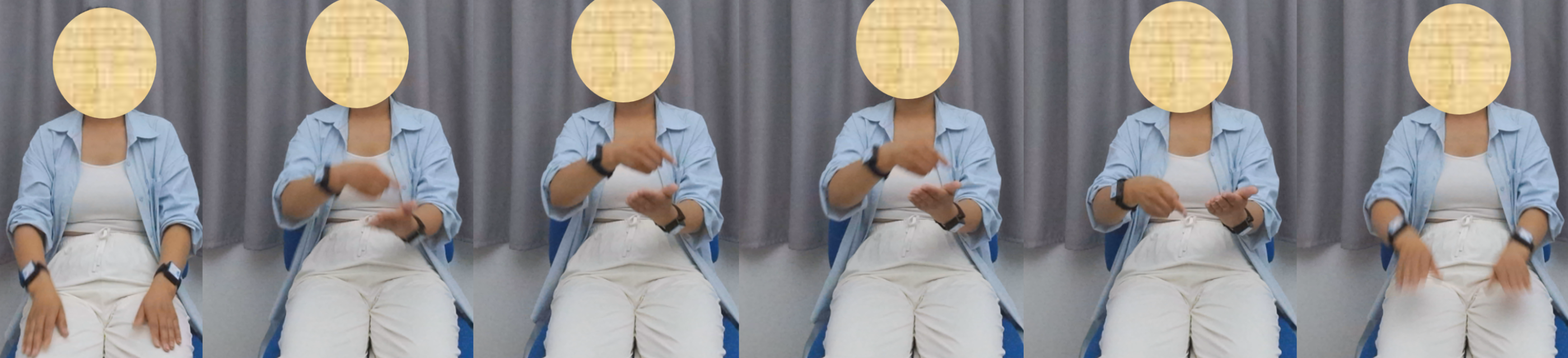}
\end{minipage}
&
\begin{minipage}[t]{\linewidth}
    \centering
    \includegraphics[width=\linewidth, height=2.2cm, keepaspectratio]{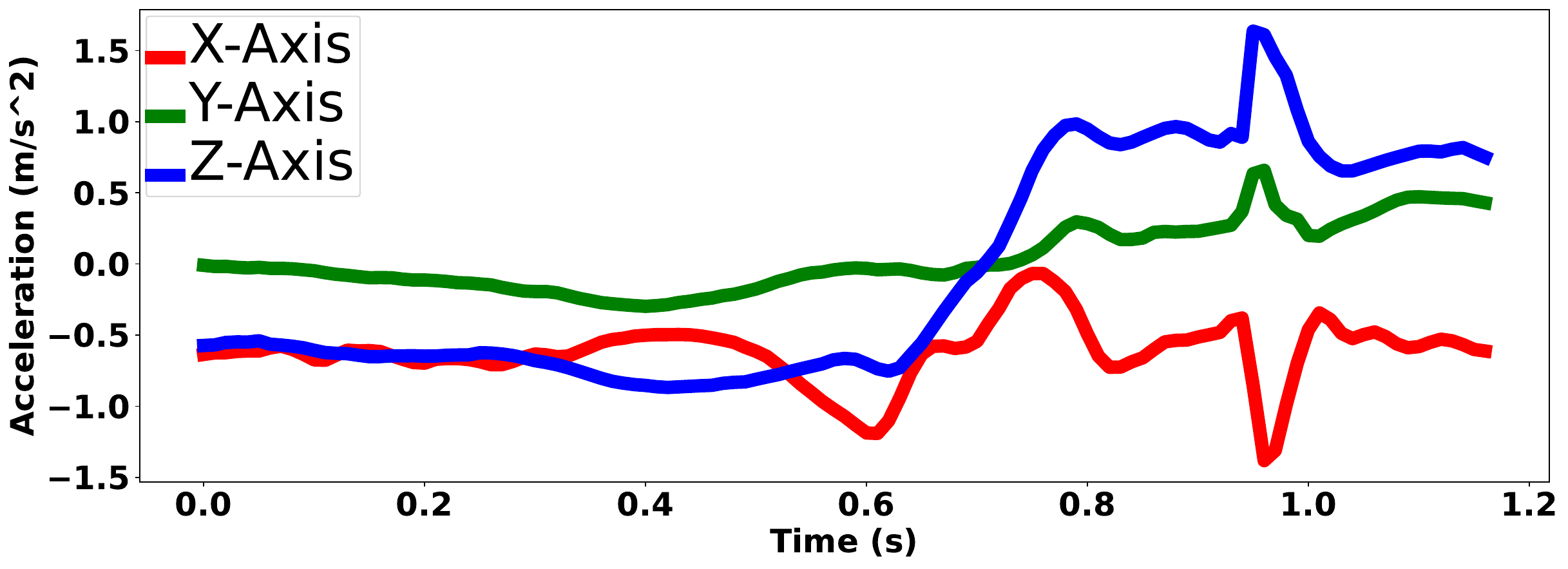}
\end{minipage}
&
\begin{minipage}[t]{\linewidth}
    \centering
    \includegraphics[width=\linewidth, height=2.2cm, keepaspectratio]{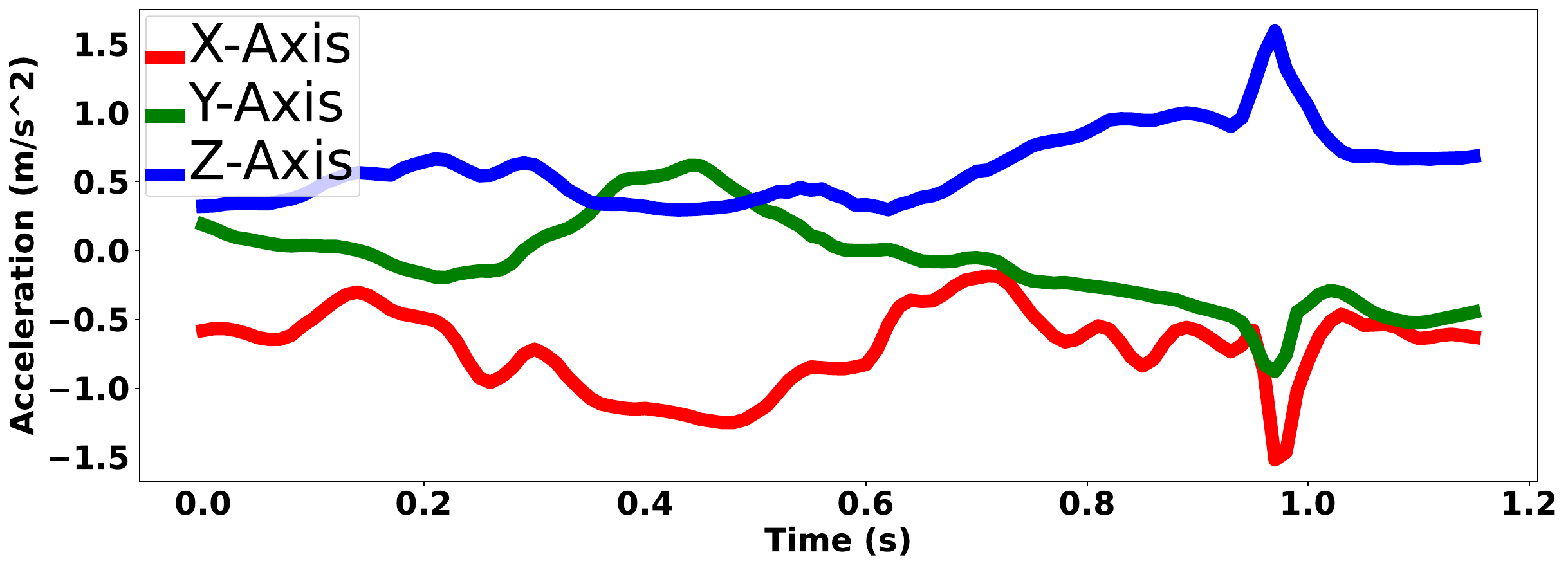}
\end{minipage}
\\ \midrule

\begin{minipage}[t]{\linewidth}
    \centering
    \includegraphics[width=\linewidth, height=2.2cm, keepaspectratio]{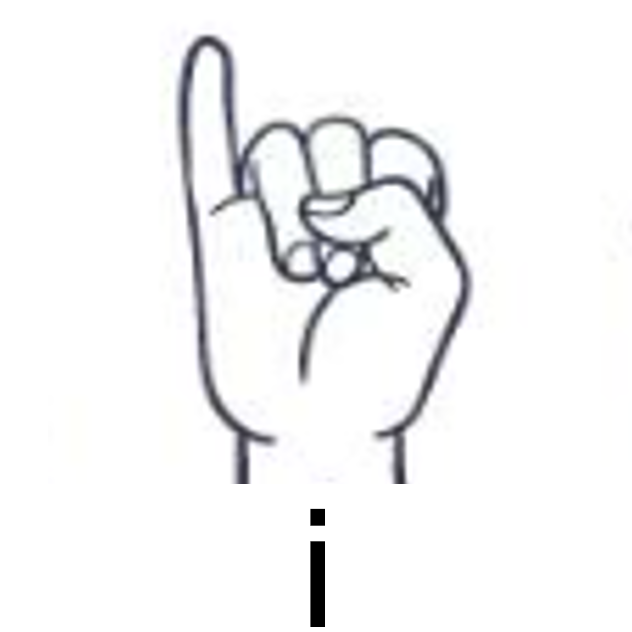}
\end{minipage}
&
\begin{minipage}[t]{\linewidth}
    \centering
    \includegraphics[width=\linewidth, height=2.2cm, keepaspectratio]{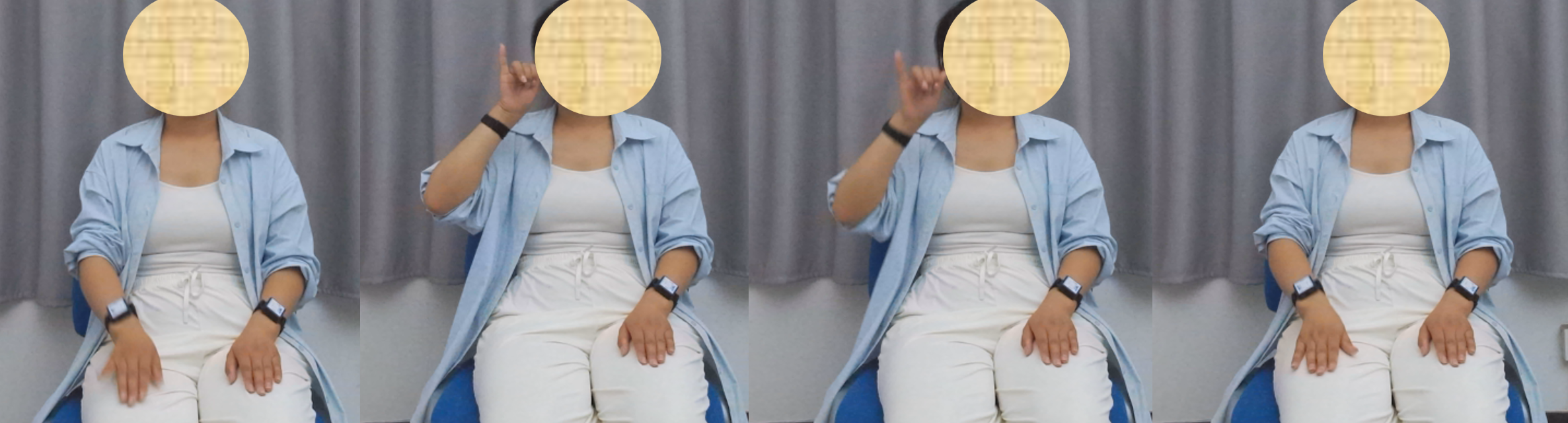}
\end{minipage}
&
\begin{minipage}[t]{\linewidth}
    \centering
    \includegraphics[width=\linewidth, height=2.2cm, keepaspectratio]{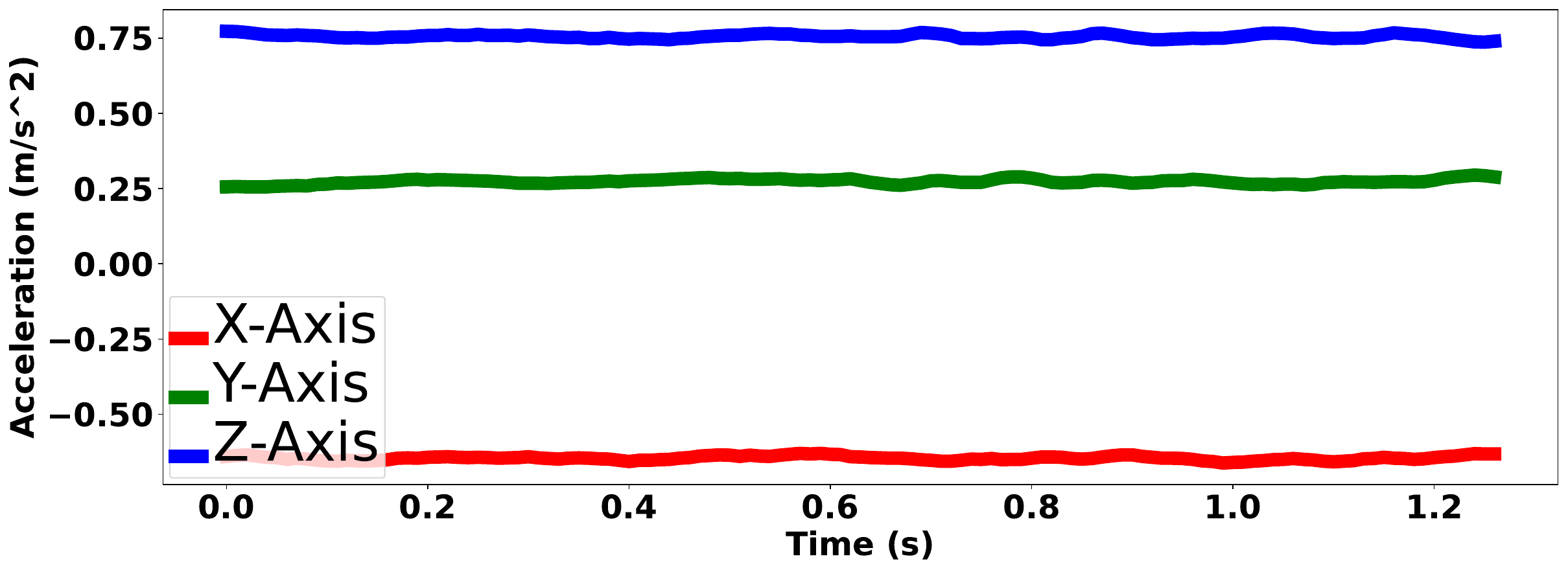}
\end{minipage}
&
\begin{minipage}[t]{\linewidth}
    \centering
    \includegraphics[width=\linewidth, height=2.2cm, keepaspectratio]{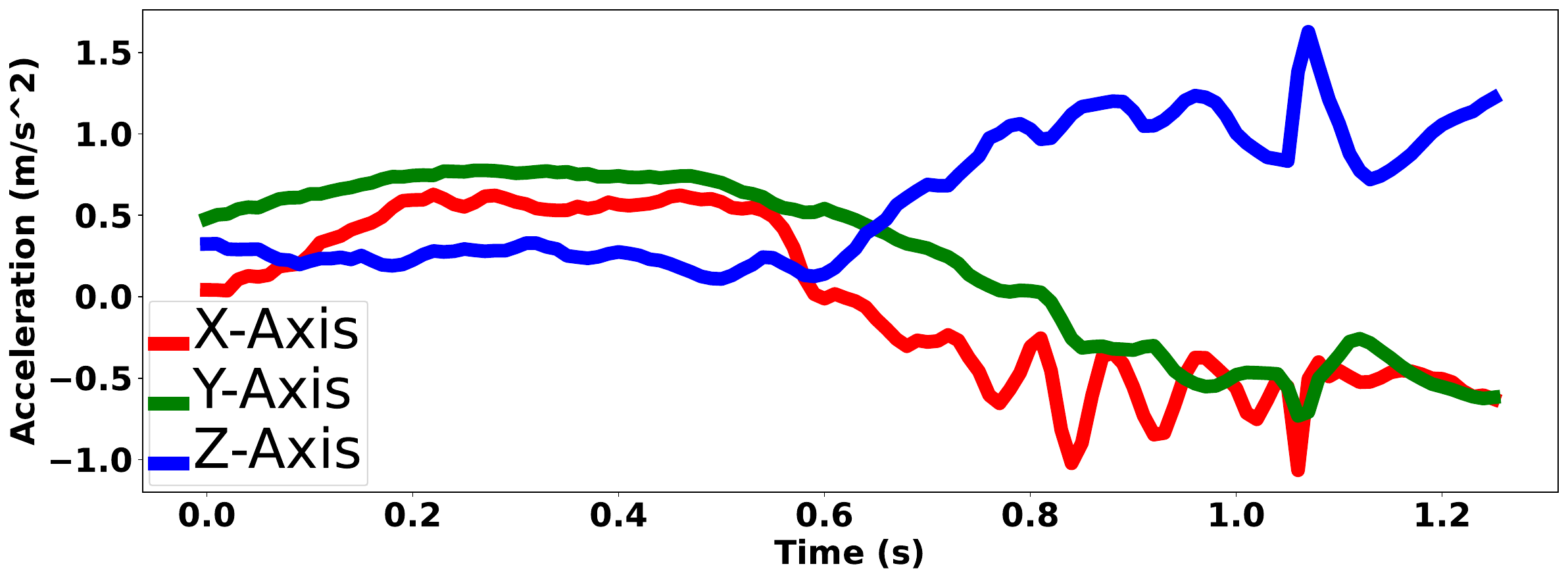}
\end{minipage}
\\
\begin{minipage}[t]{\linewidth}
    \centering
    \includegraphics[width=\linewidth, height=2.2cm, keepaspectratio]{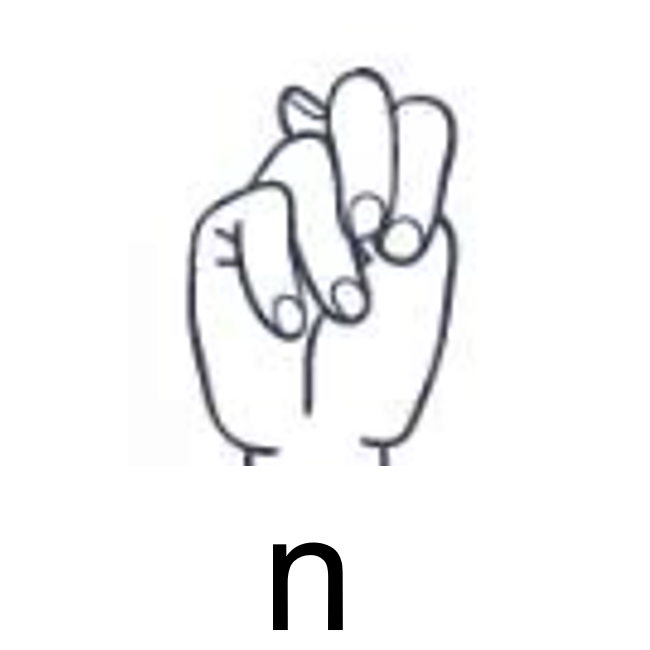}
\end{minipage}
&
\begin{minipage}[t]{\linewidth}
    \centering
    \includegraphics[width=\linewidth, height=2.2cm, keepaspectratio]{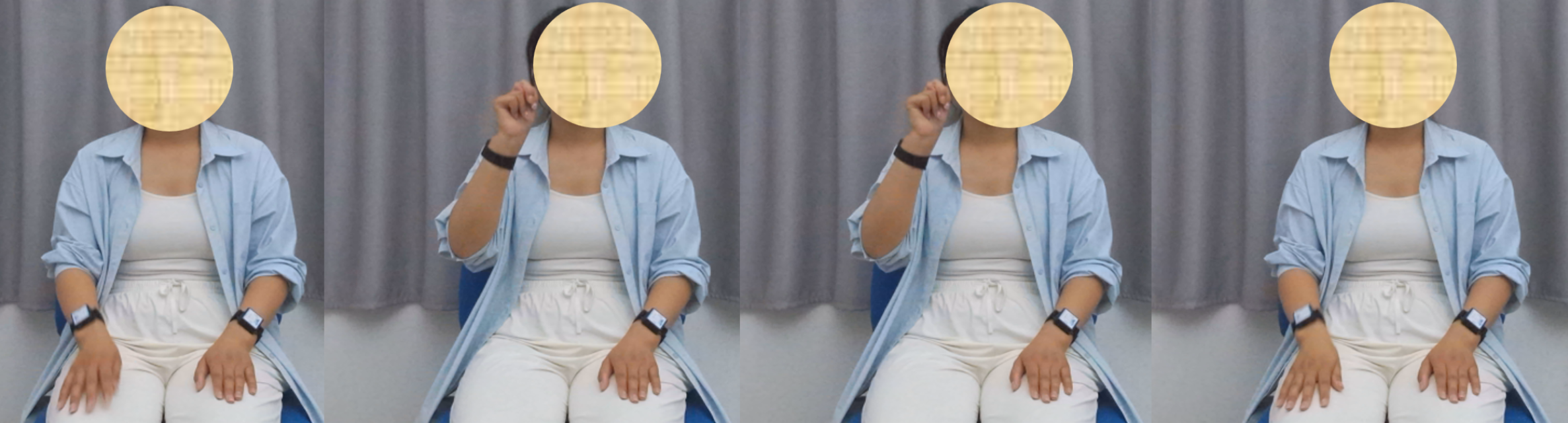}
\end{minipage}
&
\begin{minipage}[t]{\linewidth}
    \centering
    \includegraphics[width=\linewidth, height=2.2cm, keepaspectratio]{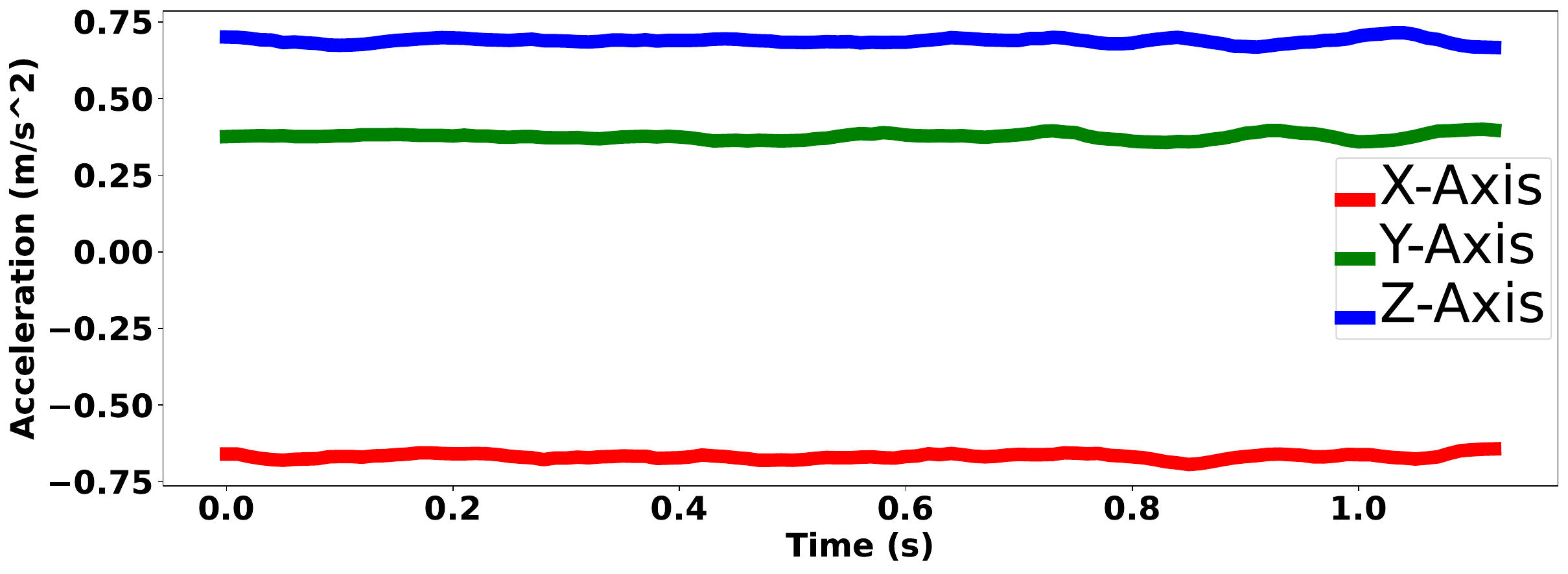}
\end{minipage}
&
\begin{minipage}[t]{\linewidth}
    \centering
    \includegraphics[width=\linewidth, height=2.2cm, keepaspectratio]{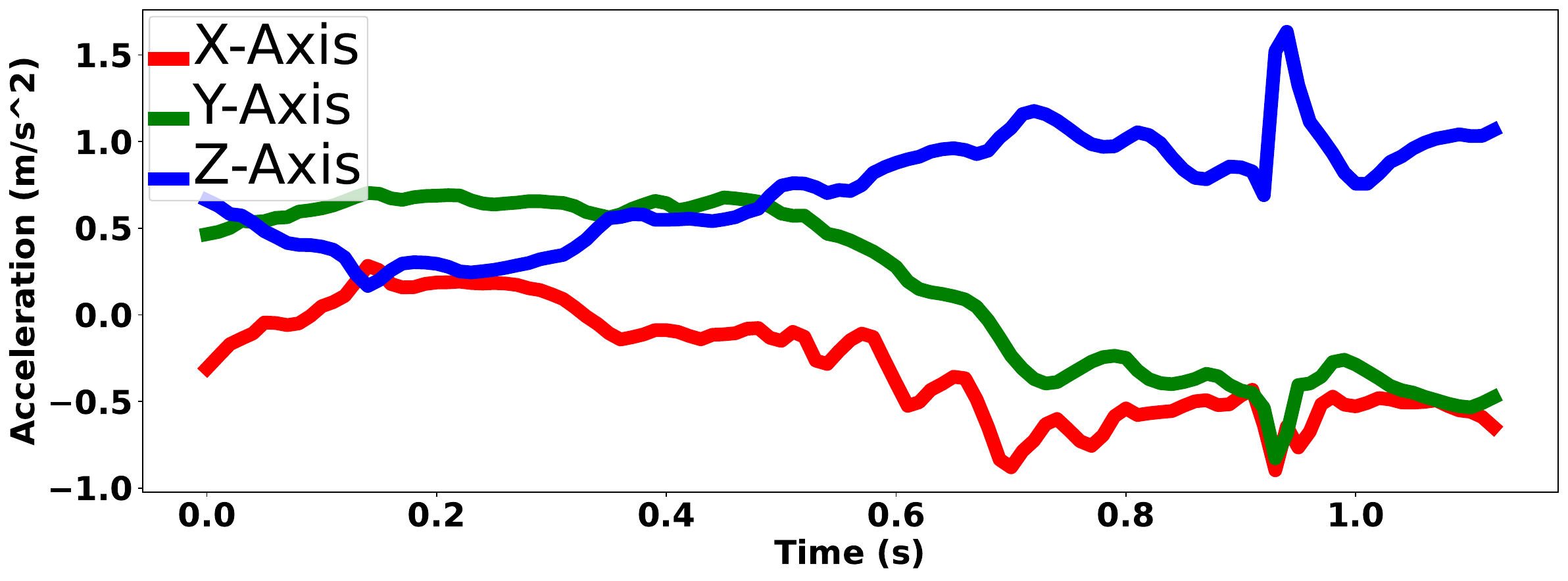}
\end{minipage}
\\ \bottomrule
\end{tabular}
\caption{\added{Examples of ambiguous words visualized in RGB sequences and IMU signals.}}
\label{tab:rgb-imu}

\end{figure*}

\subsection{Multi-Modality Evaluation Results}
\label{sec:multi_validation}

As I3D consistently achieves superior accuracy across all modalities in single-modality evaluations, we adopt it as the backbone feature extractor for all modalities in the multimodal evaluation. The 3D convolutional architecture of I3D effectively captures joint spatial and temporal cues, making it particularly suitable for video-based recognition tasks.
Table~\ref{tab:multi_modality_fusion_results} summarizes the results for all pairwise modality combinations as well as the integration of all four modalities under both UD and UI protocols. From the results, it is evident that multimodal fusion does not always lead to higher accuracy compared with single-modality baselines. For example, fusing RGB with IMU or RGB with mmWave fails to surpass the performance of RGB alone, indicating that simple fusion strategies may not fully exploit cross-modal complementarity.
Nevertheless, combinations such as RGB–Depth and IMU–mmWave yield clear performance improvements over their individual modalities, validating the benefits of multimodal integration. These findings underscore both the promise and the challenges of multimodal fusion for ISLR, suggesting that future research should explore more adaptive and semantically aligned fusion mechanisms to better harness complementary signals across heterogeneous sensing modalities.

\begin{table*}[htb]
  \centering
  \caption{Multi-modal I3D fusion results on \systemname{} under \textbf{UI} and \textbf{UD} protocols. Each entry reports Top-1 / Top-3 / Top-5 accuracies (\%).}
  \label{tab:multi_modality_fusion_results}
  \begin{tabular}{l|ccc|ccc}
    \toprule
    \multirow{2}{*}{\textbf{Modality Combination}} &
    \multicolumn{3}{c|}{\textbf{User-Independent (UI)}} & 
    \multicolumn{3}{c}{\textbf{User-Dependent (UD)}} \\
    \cmidrule(lr){2-4} \cmidrule(lr){5-7}
     & \textbf{Top-1} & \textbf{Top-3} & \textbf{Top-5} & 
       \textbf{Top-1} & \textbf{Top-3} & \textbf{Top-5} \\
    \midrule
    RGB + Depth                  & 89.32 & 96.42 & 97.92 & 94.45 & 98.81 & 99.35 \\
    RGB + IMU                    & 85.25 & 94.39 & 96.45 & 92.03 & 97.82 & 98.65 \\
    RGB + Radar                  & 83.40 & 93.30 & 95.61 & 91.54 & 97.50 & 98.48 \\
    Depth + IMU                  & 75.50 & 88.19 & 91.80 & 82.13 & 92.67 & 95.32 \\
    Depth + mmWave               & 74.63 & 88.78 & 92.39 & 82.06 & 93.12 & 95.76 \\
    IMU + mmWave                 & 75.82 & 87.18 & 90.91 & 81.35 & 91.07 & 94.22 \\
    RGB + Depth + IMU + mmWave   & 87.31 & 95.73 & 97.45 & 93.12 & 98.39 & 99.08 \\
    \bottomrule
  \end{tabular}
\end{table*}

\section{Limitation and Discussion}

This section discusses the main limitations of the current \systemname{} dataset and benchmarking framework, followed by potential research directions that can further advance multimodal SLR.

\subsection{Limited Ecological Validity and Participant Diversity}

Despite the careful design of our data acquisition system and protocol, \systemname{} still faces two primary limitations related to ecological validity.
First, the current dataset includes 20 participants\added{, all of whom are novice signers,} which may not fully capture the diversity of signing styles across different regions, age groups, and signing proficiencies\added{. In particular, Deaf/HoH native signers often exhibit higher signing speed, stronger co-articulation between signs, richer non-manual markers (e.g., facial expressions), and broader stylistic and lexical variation (e.g., regional and community-specific variants) than novice learners. These factors may introduce a domain shift when models trained on novice, studio-recorded, isolated word-level signs are applied to naturalistic Deaf/HoH signing. Therefore, we position \systemname{} as a strong benchmark and public resource for learner-oriented and controlled scenarios, while acknowledging limitations for direct deployment in real-world settings. Expanding the participant pool to include Deaf/HoH signers and expert interpreters will be crucial to improving the generalization of recognition models across diverse user populations}.
Second, all recordings were conducted under controlled laboratory conditions, which, while ensuring high data quality and synchronization accuracy, limit environmental variability such as dynamic backgrounds, spontaneous signing, or multi-signer interactions\added{. 
Future work can focus on: (1) recruiting Deaf/HoH native signers as well as professional interpreters, in close consultation and collaboration with the Deaf community to ensure culturally and ethically appropriate study design; (2) collecting continuous, sentence-level signing in addition to isolated word signs to better model co-articulation; and (3) increasing environmental diversity (e.g., lighting, background, and camera--subject distance) to evaluate robustness under real-world and ``in-the-wild'' conditions}.

\subsection{Cross-Sensor Generalization and Transferability}

\added{In this work, we use a single hardware setup (Azure Kinect for depth sensing and TI IWR1443 for mmWave radar), acknowledging that this may limit the generality and transferability of our models across different devices and configurations. Recent research has increasingly targeted cross-device and cross-domain challenges through techniques such as unsupervised domain adaptation \cite{zhang2023unsupervised}, transfer learning \cite{liu2022mtranssee}, and adversarial knowledge transfer \cite{kamboj2023transfer}. However, universal generalization remains an ongoing challenge. Strategies such as self-aligned \cite{pinyoanuntapong2023gaitsada} and active domain adaptation \cite{lin2025active} further highlight the inherent complexity of these scenarios. To support future efforts, we provide detailed radar configurations and parameters to facilitate model adaptation and fine-tuning on alternative hardware.}

\subsection{Better Multimodal Fusion Design}

As demonstrated in Sec.~\ref{sec:multi_validation}, simple fusion strategies do not consistently outperform single-modality models. While certain combinations, such as IMU and mmWave, show promising complementarity, others (e.g., RGB with IMU or mmWave) yield limited gains, suggesting that naive feature concatenation or averaging cannot fully exploit cross-modal potential. \added{We attribute this behavior to the following factors: (i) the high discriminative power of RGB under clean lab conditions; (ii) the inability of non-visual modalities to provide more information than RGB, or the difficulty in extracting recognizable cues; for many classes (e.g., signs dominated by fine-grained finger articulation), IMU/mmWave provide noisy or conflicting evidence, which can perturb otherwise-correct RGB predictions (negative transfer); (iii) using global fusion weights shared across all samples cannot selectively "activate" sensors only when they are reliable (instance dependence), and logit-level fusion cannot model the richer cross-modal interactions needed for alignment and complementary reasoning. Promising directions include instance- and uncertainty-aware gating~\cite{lee2016stochastic}, conditional activation to suppress unreliable modalities~\cite{perez2018film}, and mid-level fusion with cross-attention to enable feature-level interactions and alignment~\cite{tsai2019multimodal}.} Moreover, lightweight multimodal frameworks optimized for wearable or mobile platforms could enable real-time, privacy-preserving sign recognition in practical ubiquitous computing scenarios.

\subsection{Opportunities for Cross-Modal Generation and Augmentation}

Existing large-scale SLR datasets are predominantly visual, focusing on video-based recognition and translation tasks with vocabularies of thousands of signs. In contrast, datasets incorporating non-visual modalities such as mmWave radar or IMUs remain scarce and small in scale, typically containing only tens of signs. Models trained on such limited data often fail to generalize and are insufficient for realistic sign language recognition or translation applications.
With its precisely synchronized multimodal recordings, \systemname{} enables new opportunities for cross-modal generation and data augmentation. For example, mappings can be learned between visual and radar domains to synthesize mmWave spectrograms from video inputs, thereby expanding the availability of non-visual training data\added{~\cite{deng2023midas,deng2024g}}. Such cross-modal generation strategies can facilitate few-shot or even zero-shot learning, enhancing model robustness and generalization in modalities where public datasets remain scarce.

\section{Dataset and Code Availability}
\myadded{
The code and dataset are publicly available at \url{https://github.com/happy2sumture-cloud/SIGMA-ASL}.}

\section{Conclusion}

This paper introduced \systemname{}, a large-scale multimodal dataset for isolated sign language recognition that integrates synchronized RGB-D video, mmWave radar, and IMU sensing. Unlike existing vision-centric datasets, \systemname{} provides complementary visual, radio, and kinematic information within a unified sensing framework, achieving millisecond-level temporal synchronization across modalities. We established standardized preprocessing pipelines and benchmarking protocols under both user-dependent and user-independent settings, enabling consistent evaluation of single- and multi-modality models. The experimental results present comprehensive baseline performance, offering valuable references for future studies. As a foundational resource, \systemname{} is designed to advance research in robust, privacy-preserving, and ubiquitous sign language understanding, fostering new directions in multimodal perception, cross-modal learning, and inclusive human–computer interaction.

\begin{acks}
This work is partially supported by the Natural Science Foundation of Shandong Province (Major Basic Research) Grant No.~ZR2024ZD12, the Fundamental Research Funds of Shandong University, and the Open Project Program of the State Key Laboratory of Virtual Reality Technology and Systems, Beihang University (No.~VRLAB2025C04).
\end{acks}

\bibliographystyle{ACM-Reference-Format}
\bibliography{references}

\clearpage
\appendix
\section{Language Acronyms Used in Table~\ref{tab:ISLR datasets}}
\label{app:language_acronyms}

Table~\ref{tab:ISLR datasets} includes multiple sign languages abbreviated for brevity. 
Their full names are as follows:

\begin{table}[h]
\centering
\footnotesize
\caption{List of language acronyms and their full names.}
\label{tab:language_acronyms_full}
\begin{tabular}{ll}
\toprule
\textbf{Acronym} & \textbf{Full Name (English Translation)} \\
\midrule
ASL & American Sign Language \\
BSL & British Sign Language \\
CSL & Chinese Sign Language \\
DSGS & Swiss German Sign Language \\
GSL & Greek Sign Language \\
ISL & Indian Sign Language \\
LSR & Lengua de Señas Argentina (Argentinian Sign Language) \\
LSE & Lengua de Signos Española (Spanish Sign Language) \\
LSFB & Langue des Signes de Belgique Francophone (Belgian French Sign Language) \\
TSL & Turkish Sign Language \\
Auslan & Australian Sign Language \\
PSL & Persian Sign Language \\
\bottomrule
\end{tabular}
\end{table}

\section{Vocabulary List of the SIGMA-ASL Dataset}
\label{app:vocabulary_list}

\begin{table*}[h]
\centering
\footnotesize
\setlength{\tabcolsep}{5pt}
\renewcommand{\arraystretch}{1.1}
\caption{Complete vocabulary list of the SIGMA-ASL dataset (160 signs).
Words marked with numerical suffixes (e.g., “word\textsuperscript{1}”) indicate variant signs of the same word.}
\label{tab:vocab_list}
\begin{tabularx}{\textwidth}{XXXXXXXX}
\toprule
mom & dad & boy & girl & marriage & husband & wife & brother \\
sister & grandma & grandpa & aunt & uncle & baby & single & divorced \\
home & work & school & store & store\textsuperscript{1} & church & come & go \\
car & in & out & with & with\textsuperscript{1} & day & day\textsuperscript{1} & day\textsuperscript{2} \\
night & every night & week & month & every month & year & will & before \\
today & today\textsuperscript{1} & finish & hot & cold & pizza & pizza\textsuperscript{1} & milk \\
hamburger & hot dog & egg & apple & cheese & drink & spoon & fork \\
cup & cereal & water & candy & cookie & hungry & shirt & pants \\
socks & shoes & coat & underwear & wash & hurt & hurt\textsuperscript{1} & bathroom \\
brush teeth & sleep & nice & happy & angry & sad & sorry & cry \\
like & good & good\textsuperscript{1} & bad & love & please & excuse & thank you \\
help & who & what & what\textsuperscript{1} & when & where & why & how \\
stop & big & tall & full & more & blue & green & yellow \\
red & brown & orange & gold & silver & dollars & cent & cost \\
cat & dog & bird & horse & cow & sheep & pig & bug \\
here & child & welcome & same & one & two & three & four \\
five & six & seven & eight & nine & ten & a & b \\
c & d & e & f & g & h & i & j \\
k & l & m & n & o & p & q & r \\
s & t & u & v & w & x & y & z \\
\bottomrule
\end{tabularx}

\end{table*}

The SIGMA-ASL dataset contains 160 isolated American Sign Language (ASL) words, including 103 common vocabulary items, 10 digits (0–9), 26 alphabet letters (A–Z), and 21 gesture variants (e.g., \textit{single}, \textit{apple}, \textit{sorry}). 
Each variant represents a perceptually and kinematically distinct motion pattern. 
Table~\ref{tab:vocab_list} provides the full list of vocabulary items used in the dataset.

\section{Data Distribution and Balance Analysis}
\label{app:data_dist}

\myadded{
To assess potential imbalance, we report the distribution of clip counts across (i) classes and (ii) participants. Table~\ref{tab:dist_summary} shows low variability at both levels (CV $<0.1$), indicating a well-balanced dataset. Table~\ref{tab:discard_per_user} reports per-participant discarded clips after filtering, with consistently low discard rates ($\approx$1.5\%--2.3\%).} (\textcolor{blue}{p28})

\begin{table}[t]
\centering
\caption{Summary statistics of clip-count distribution across classes and participants.}
\label{tab:dist_summary}
\begin{tabular}{lrrrrrrrr}
\hline
Level & \#Groups & Total & Min & Max & Mean & Std & CV & Max/Min \\
\hline
Per-class (exp) & 160 & 93{,}545 & 495 & 706 & 584.66 & 50.41 & 0.086 & 1.43 \\
Per-participant & 20  & 93{,}545 & 4{,}132 & 5{,}766 & 4{,}677.25 & 368.93 & 0.079 & 1.40 \\
\hline
\end{tabular}
\end{table}

\begin{table}[t]
\centering
\caption{Discarded clips per participant(before vs. after filtering).}
\label{tab:discard_per_user}
\begin{tabular}{lrrrr}
\hline
Participant & Before & After & Discarded & Discard rate (\%) \\
\hline
user\_1  & 4668 & 4576 & 92  & 1.97 \\
user\_2  & 4393 & 4292 & 101 & 2.30 \\
user\_3  & 4202 & 4132 & 70  & 1.67 \\
user\_4  & 5077 & 4959 & 118 & 2.33 \\
user\_5  & 4569 & 4480 & 89  & 1.95 \\
user\_6  & 4443 & 4348 & 95  & 2.14 \\
user\_7  & 4625 & 4556 & 69  & 1.49 \\
user\_8  & 4513 & 4428 & 85  & 1.88 \\
user\_9  & 4781 & 4675 & 106 & 2.22 \\
user\_10 & 4988 & 4914 & 74  & 1.48 \\
user\_11 & 4764 & 4682 & 82  & 1.72 \\
user\_12 & 5450 & 5357 & 93  & 1.71 \\
user\_13 & 5859 & 5766 & 93  & 1.59 \\
user\_14 & 4775 & 4697 & 78  & 1.63 \\
user\_15 & 4582 & 4489 & 93  & 2.03 \\
user\_16 & 4434 & 4342 & 92  & 2.08 \\
user\_17 & 4992 & 4908 & 84  & 1.68 \\
user\_18 & 4928 & 4829 & 99  & 2.01 \\
user\_19 & 4599 & 4503 & 96  & 2.09 \\
user\_20 & 4700 & 4612 & 88  & 1.87 \\
\hline
\end{tabular}
\end{table}

\section{Mathematical Details of Multimodal Signal Preprocessing}
\label{app:preprocess_formulas}
This appendix provides the complete mathematical formulations of the mmWave radar
Range--Doppler Map (RDM) generation and the IMU spectrogram construction
used in the \systemname{} dataset.

\subsection{mmWave Range--Doppler Processing}
\label{app:rdm_formulas}
For each virtual channel formed by the $t$-th transmit antenna and the $u$-th receive antenna,
a Hann window $w_{\mathrm{r}}[s]$ is applied along the fast-time (ADC sample) dimension.
An $S$-point FFT is then performed to obtain the range-domain representation:
\begin{equation}
\mathcal{F}_{\mathrm{R}}(\mathbf{x}_{t,u}) =
\sum_{s=0}^{S-1}
w_{\mathrm{r}}[s]\,
\mathbf{x}_{t,u}[s]\,
e^{-j2\pi ks/S}.
\end{equation}

Static clutter is suppressed by subtracting the mean across the slow-time (chirp) dimension,
which removes near-zero Doppler components:
\begin{equation}
\mathrm{MTI}(y[l]) =
y[l] - \frac{1}{L}\sum_{l=0}^{L-1} y[l].
\end{equation}

After clutter removal, a Hann window is applied along the slow-time dimension,
followed by an $L$-point Doppler FFT:
\begin{equation}
\mathcal{F}_{\mathrm{D}}(z) =
\sum_{l=0}^{L-1}
w_{\mathrm{d}}[l]\,
z[l]\,
e^{-j2\pi ml/L}.
\end{equation}

The Doppler spectra from all $N_{\mathrm{tx}} \times N_{\mathrm{rx}}$ virtual channels
are coherently integrated to form the final Range--Doppler Map (RDM):
\begin{equation}
\mathrm{RDM}(r,v) =
20\log_{10}\!\left(
\left|
\sum_{t=1}^{N_{\mathrm{tx}}}
\sum_{u=1}^{N_{\mathrm{rx}}}
\mathcal{F}_{\mathrm{D}}
\big(
\mathrm{MTI}(
\mathcal{F}_{\mathrm{R}}(w_{\mathrm{r}} \odot \mathbf{x}_{t,u})
)
\big)
\right|
+ \varepsilon
\right),
\end{equation}
where $\varepsilon=10^{-10}$ is a small constant added for numerical stability.

\subsection{IMU Spectrogram Computation}
\label{app:imu_formulas}
Let $x_c(t)$ denote the time-domain signal of the $c$-th IMU channel.
After temporal alignment and resampling, each channel is normalized using per-channel
z-score normalization:
\begin{equation}
\hat{x}_c(t) =
\frac{x_c(t) - \mu_c}{\sigma_c},
\end{equation}
where $\mu_c$ and $\sigma_c$ denote the mean and standard deviation of channel $c$
computed within each clip.

The short-time Fourier transform (STFT) of the normalized IMU signal is defined as:
\begin{equation}
S_c(f,\tau) =
\sum_{t=-\infty}^{\infty}
\hat{x}_c(t)\,
w(t-\tau)\,
e^{-j2\pi f t},
\end{equation}
where $w(t-\tau)$ is a Hann window centered at time $\tau$.
The magnitude spectra are further log-compressed to suppress dynamic range variations:
\begin{equation}
\tilde{S}_c(f,\tau) =
\log\big(|S_c(f,\tau)| + \epsilon\big),
\end{equation}
where $\epsilon$ is a small constant for numerical stability.

Finally, the spectrograms from all IMU channels are stacked along the channel dimension
to form the IMU spectral representation:
\begin{equation}
\mathbf{X}_{\mathrm{imu}}^{\mathrm{spec}}
\in \mathbb{R}^{T \times F \times C},
\end{equation}
where $T$ is the number of temporal frames,
$F$ is the frequency resolution,
and $C=12$ is the number of IMU channels.

\subsection{Coefficient of Variation (CV)}
\label{app:cv_definition}

The Coefficient of Variation (CV) is a statistical measure used to assess the relative variability of a dataset. It is defined as the ratio of the standard deviation ($\sigma$) to the mean ($\mu$) of the data:

\begin{equation}
\text{CV} = \frac{\sigma}{\mu},
\end{equation}
where $\sigma$ is the standard deviation and $\mu$ is the mean of the dataset.

A lower CV value indicates less variability relative to the mean, suggesting that the data is more consistent. Conversely, a higher CV indicates greater variability relative to the mean. In our case, we compute the CV for both per-class clip counts and per-participant counts to assess the balance of the dataset. A low CV for both metrics (CV = 0.086 for per-class counts and CV = 0.079 for per-participant counts) suggests that the dataset is reasonably balanced, with no severe imbalances across classes or participants.

\subsection{Top-$k$ Accuracy}
\label{app:eval_formulas}
Top-$k$ Accuracy is defined as:
\begin{equation}
\text{Top-}k \ \text{Accuracy} = \frac{1}{N} \sum_{i=1}^{N} \mathbb{I}\!\left(y_i \in \hat{Y}_i^{(k)}\right),
\label{eq:topk}
\end{equation}
where $N$ is the number of test samples, $y_i$ is the ground-truth label for the $i$-th sample, $\hat{Y}_i^{(k)}$ represents the set of top-$k$ predicted labels for that sample, and $\mathbb{I}(\cdot)$ is an indicator function that equals 1 when the condition holds and 0 otherwise.

\subsection{De-identified Informed Consent Template}

The de-identified informed consent template is presented on the following page.

\includepdf[pages=-]{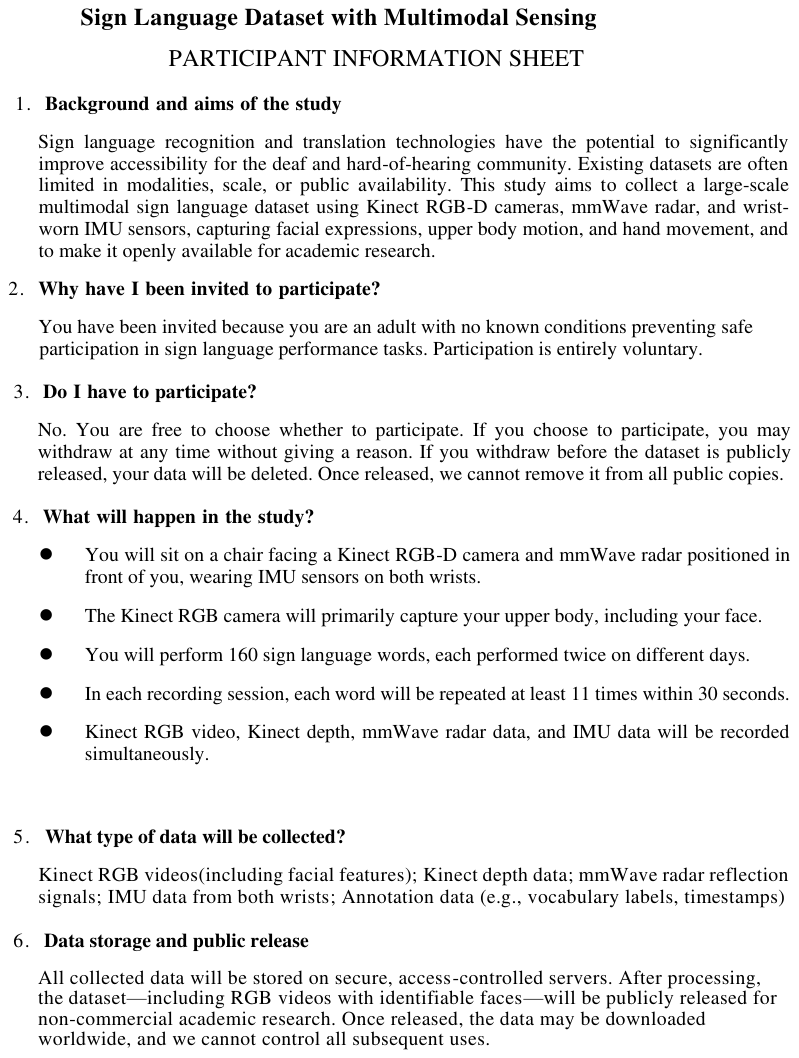}

\end{document}